\documentclass{aa}

\usepackage[nolist,nohyperlinks]{acronym}

\usepackage{siunitx}
\usepackage{microtype}

\usepackage{enumitem}

\usepackage{placeins}

\usepackage{widetext}

\usepackage{silence}
\usepackage[colorlinks, allcolors=blue, breaklinks=true]{hyperref}
\usepackage{cleveref}

\usepackage{xcolor}

\usepackage{amsmath}

\usepackage{changepage}

\usepackage{tabularx}

\usepackage{tikz}
\usetikzlibrary{shapes,arrows,fit,positioning,calc}

\let\endlongtable\relax

\usepackage[rightcaption, ragged]{sidecap}
\sidecaptionvpos{figure}{t}

\usepackage{tabularx}

\usepackage{txfonts}

\usepackage[encapsulated]{CJK}
\usepackage{ucs}

\newcommand{\orcid}[1]{\protect\href{https://orcid.org/#1}{\protect\includegraphics[width=8pt]{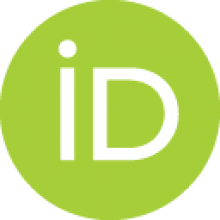}}}%%

\usepackage{xspace}

\def\M87{M87$^*$\xspace}
\def\m87{M87$^*$\xspace}
\def\sgra{Sgr~A$^*$\xspace}
\def\3C279{3C\,279\xspace}
\def\3c279{3C\,279\xspace}
\def\NRAO530{NRAO\,530\xspace}
\def\nrao530{NRAO\,530\xspace}
\def\J1924{J1924-2914\xspace}
\def\j1924{j1924-2914\xspace}

\def\symba{\mbox{\textsc{Symba}}\xspace}
\def\rpicard{\mbox{\textsc{Rpicard}}\xspace}
\def\meqsilhouette{\mbox{\textsc{MeqSilhouette}}\xspace}
\def\tensorflow{\mbox{\textsc{TensorFlow}}\xspace}

\def\atm{\mbox{\textsc{atm}}\xspace}

\begin{document}

\begin{acronym}
\acro{sn}[S/N]{signal-to-noise ratio}
\acro{vlbi}[VLBI]{very long baseline interferometry}
\acroplural{vlbi}[VLBI]{Very-long-baseline interferometry}
\acro{grmhd}[GRMHD]{general relativistic magnetohydrodynamics}
\acro{grrt}[GRRT]{general relativistic ray-tracing}
\acro{mhd}[MHD]{magnetohydrodynamics}
\acro{as}[as]{arcseconds}
\acro{agn}[AGNs]{Active Galactic Nuclei}
\acro{llagn}[LLAGN]{low-luminosity AGN}
\acro{riaf}[RIAF]{radiatively inefficient accretion flow}
\acro{cena}[Cen~A]{Centaurus~A}
\acroplural{cena}[Cen~A, NGC~5128]{Centaurus~A}
\acro{sgra}[Sgr\,A*]{Sagittarius~A*}
\acro{m87}[\m87]{Messier 87*}
\acro{agn}[AGNs]{active galactic nuclei}
\acro{eht}[EHT]{Event Horizon Telescope}
\acro{bhc}[BHC]{\href{https://blackholecam.org}{BlackHoleCam}}
\acro{tanami}[TANAMI]{Tracking Active Galactic Nuclei with Austral Milliarcsecond Interferometry}
\acro{smbh}[SMBH]{supermassive black hole}
\acroplural{smbh}[SMBHs]{supermassive black holes}
\acro{aa}[ALMA]{Atacama Large Millimeter/submillimeter Array}
\acro{ap}[APEX]{Atacama Pathfinder Experiment}
\acro{pv}[PV]{IRAM~30\,m Telescope}
\acro{jc}[JCMT]{James Clerk Maxwell Telescope}
\acro{lm}[LMT]{Large Millimeter Telescope Alfonso Serrano}
\acro{sp}[SPT]{South Pole Telescope}
\acro{sm}[SMA]{Submillimeter Array}
\acro{az}[SMT]{Submillimeter Telescope}
\acro{c}[c]{speed of light}
\acro{pc}[pc]{parsec}
\acro{gr}[GR]{general relativity}
\acro{aips}[\textsc{aips}]{\href{http://www.aips.nrao.edu}{Astronomical Image Processing System}}
\acro{casa}[\textsc{casa}]{\href{https://casa.nrao.edu}{Common Astronomy Software Applications}}
\acro{symba}[\textsc{Symba}]{\href{https://bitbucket.org/M_Janssen/symba}{SYnthetic Measurement creator for long Baseline Arrays}}
\acro{rpicard}[\textsc{Rpicard}]{\href{https://bitbucket.org/M_Janssen/picard}{Radboud PIpeline for the Calibration of high Angular Resolution Data}}
\acro{hops}[\textsc{hops}]{\href{https://www.haystack.mit.edu/tech/vlbi/hops.html}{Haystack Observatory Postprocessing System}}
\acro{hbt}[HBT]{Hanbury Brown and Twiss}
\acro{tov}[TOV]{Tolman-Oppenheimer-Volkoff}
\acro{mri}[MRI]{magnetorotational instability}
\acro{adaf}[ADAF]{advection-dominated accretion flow}
\acro{adios}[ADIOS]{adiabatic inflow-outflow solution}
\acro{cdaf}[CDAF]{convection-dominated accretion flow}
\acro{bz}[BZ]{Blandford-Znajek}
\acro{bp}[BP]{Blandford-Payne}
\acro{em}[EM]{electromagnetic}
\acro{blr}[BLR]{broad-line region}
\acro{nlr}[NLR]{narrow-line region}
\acro{ism}[ISM]{interstellar medium}
\acro{edf}[eDF]{electron distribution function}
\acro{pic}[PIC]{particle-in-cell}
\acro{sane}[SANE]{standard and normal evolution}
\acro{mad}[MAD]{magnetically arrested disk}
\acro{jive}[JIVE]{\href{http://www.jive.eu}{Joint Institute for VLBI ERIC}}
\acro{nrao}[NRAO]{\href{https://www.nrao.edu}{National Radio Astronomy Observatory}}

\acro{muas}[$\mu$as]{microarcseconds}
\acroplural{agn}[AGNs]{Active galactic nuclei}
\acro{jy}[Jy]{Jansky}
\acro{pa}[PA]{position angle}
\acro{srmhd}[SRMHD]{special relativistic magnetohydrodynamics}
\end{acronym}

\title{Deep learning inference with the Event Horizon Telescope}
\subtitle{I. Calibration improvements and a comprehensive synthetic data library}

   \author{M. Janssen\orcid{0000-0001-8685-6544}\inst{1,2}
          \and 
          C.-k. Chan\orcid{0000-0001-6337-6126}\inst{3,4,5}
          \and
          J. Davelaar\orcid{0000-0002-2685-2434}\inst{6,7}
          \and
          I. Natarajan\orcid{0000-0001-8242-4373}\inst{8,9}
          \and
          H. Olivares\orcid{0000-0001-6833-7580}\inst{10}
          \and
          B.~Ripperda\orcid{0000-0002-7301-3908}\inst{11,12,13,14}
          \and
          J.~R\"oder\orcid{0000-0002-2426-927X}\inst{2}
          \and
          M.~Rynge\orcid{0000-0002-1779-7189}\inst{15}
          \and
          M.~Wielgus\orcid{0000-0002-8635-4242}\inst{16}
          }
    \institute{Department of Astrophysics, Institute for Mathematics, Astrophysics and Particle Physics (IMAPP), Radboud University, P.O. Box 9010, 6500 GL Nijmegen, The Netherlands\\
    \email{M.Janssen@astro.ru.nl}
   \and Max-Planck-Institut f\"ur Radioastronomie, Auf dem H\"ugel 69, D-53121 Bonn, Germany
            \and
            Steward Observatory and Department of Astronomy, University of Arizona, 933 N. Cherry Ave., Tucson, AZ 85721, USA
            \and
            Data Science Institute, University of Arizona, 1230 N. Cherry Ave., Tucson, AZ 85721, USA
            \and
            Program in Applied Mathematics, University of Arizona, 617 N. Santa Rita Ave., Tucson, AZ 85721, USA
            \and
            Department of Astrophysical Sciences, Peyton Hall, Princeton University, Princeton, NJ 08544, USA
            \and
            NASA Hubble Fellowship Program, Einstein Fellow
            \and
            Center for Astrophysics | Harvard \& Smithsonian, 60 Garden Street, Cambridge, MA 02138, USA
            \and
            Black Hole Initiative at Harvard University, 20 Garden Street, Cambridge, MA 02138, USA
            \and
            Departamento de Matem\'{a}tica da Universidade de Aveiro and Centre for Research and Development in Mathematics and Applications (CIDMA), Campus de Santiago, 3810-193 Aveiro, Portugal
            \and
            Canadian Institute for Theoretical Astrophysics, University of Toronto, Toronto, ON, Canada M5S 3H8
            \and
            David A. Dunlap Department of Astronomy, University of Toronto, 50 St. George Street, Toronto, ON M5S 3H4
            \and 
            Department of Physics, University of Toronto, 60 St. George Street, Toronto, ON M5S 1A7
            \and 
            Perimeter Institute for Theoretical Physics, Waterloo, Ontario N2L 2Y5, Canada
            \and
            University of Southern California - Information Sciences Institute, 4676 Admiralty Way, Suite 1001, Marina del Rey, CA 90292, USA
            \and Instituto de Astrofísica de Andalucía-CSIC, Glorieta de la Astronomía s/n, E-18008 Granada, Spain}

   \date{Received TBD; accepted TBD}

\nocite{eht-paperI}
\nocite{eht-paperII}
\nocite{eht-paperIII}
\nocite{eht-paperIV}
\nocite{eht-paperV}
\nocite{eht-paperVI}
\nocite{eht-m87-paper-vii}
\nocite{eht-SgrAi}
\nocite{eht-SgrAii}
\nocite{eht-SgrAiii}
\nocite{eht-SgrAiv}
\nocite{eht-SgrAv}
\nocite{eht-SgrAvi}

\abstract
{In a series of publications, we describe a comprehensive comparison of Event Horizon Telescope (EHT) data with theoretical models of the observed Sagittarius~A* (\sgra) and Messier 87* (\m87) horizon-scale sources.}
{In this first article, we report on improvements made to our observational data reduction pipeline and present the generation of observables derived from the EHT models. We make use of ray-traced general relativistic magnetohydrodynamic simulations that are based on different black hole spacetime metrics and accretion physics parameters. These broad classes of models provide a good representation of the primary targets observed by the EHT.}
{We describe how we combined multiple frequency bands and polarization channels of the observational data to improve our fringe-finding sensitivity and stabilization of atmospheric phase fluctuations. To generate realistic synthetic data from our models, we took the signal path as well as the calibration process, and thereby the aforementioned improvements, into account. We could thus produce synthetic visibilities akin to calibrated EHT data and identify salient features for the discrimination of model parameters.
}
{We have produced a library consisting of an unparalleled 962,000 synthetic \sgra and \m87 datasets. In terms of baseline coverage and noise properties, the library encompasses 2017 EHT measurements as well as future observations with an extended telescope array.} %852000 + e17e11lo KN + ATM KN + e17c07 D + ngEHT D
{We differentiate between robust visibility data products related to model features and data products that are strongly affected by data corruption effects. Parameter inference is mostly limited by intrinsic model variability, which highlights the importance of long-term monitoring observations with the EHT. In later papers in this series, we will show how a Bayesian neural network trained on our synthetic data is capable of dealing with the model variability and extracting physical parameters from EHT observations. With our calibration improvements, our newly reduced EHT datasets have a considerably better quality compared to previously analyzed data.}

\maketitle

\ActivateWarningFilters[hreflink]
\section{Introduction} 

Powered by accretion onto \acp{smbh}, \ac{agn} belong to the most luminous persistent sources in the known Universe \citep{AGN1,AGN2}.
With the \ac{eht} \ac{vlbi} array, we can resolve the innermost region of \ac{agn} and study accretion onto compact objects, plasma physics, jet launching, and gravity in the strong field regime \citep{eht-paperI, eht-paperII, 2020Kim, 2020Psaltis, eht-m87-paper-vii, 2021Kocherlakota, 2021EHTCENA, eht-SgrAi, eht-SgrAvi, 2022Issaoun, 2023Jorstad, Paraschos2024,Baczko2024}.
However, interpreting EHT data poses significant challenges, such as the inherent variability of black hole accretion flows, atmospheric noise, and instrument-related data corruption effects.
Addressing these issues requires both methodological advances and high-fidelity simulations of synthetic data.

Currently, \ac{grmhd} simulations \citep[e.g.,][]{2022Mizuno} are the most complete class of models available for a self-consistent description of \ac{agn}' central activity.
\ac{grmhd}s naturally produce jets and, without (substantial) radiative cooling, a~geometrically thick accretion flow.
For parameters matching the \ac{sgra} and \ac{m87} sources, \ac{grrt} calculations based on \ac{grmhd} simulations predict \mbox{(sub-)}millimeter synchrotron emission to be produced in an (mostly) optically thin plasma surrounding the black hole \citep{eht-paperV, eht-SgrAv}. From \ac{grmhd}-\ac{grrt} models, full Stokes horizon-scale movies of the disk plus jet emission of \ac{sgra} and \ac{m87} have been made.
The geometrically thick and optically thin accretion scenario with accompanying jets is suitable for low-powered \ac{agn} \citep[e.g.,][and references therein]{2008Ho}.
The activity of luminous \ac{agn} and X-ray binaries can also be simulated with \ac{grmhd}-\ac{grrt} models \citep[e.g.,][]{2021Dexter, 2021Liska,Wielgus2022}, but unlike low accretion rate sources, they generally require radiative cooling to be self-consistent.
For the \ac{eht}, \ac{grmhd}-\ac{grrt} simulations are used for the validation of image reconstructions \citep{eht-paperIV, eht-SgrAiii} and tests of \ac{gr} together with the inference of MHD, accretion, and black hole parameters \citep{eht-paperV, eht-paperVI, 2020Psaltis, 2021Kocherlakota, eht-SgrAv, eht-SgrAvi}.

In this first paper of a series, we present a comprehensive synthetic data library that is based on a wide range of \ac{grmhd}-\ac{grrt} models.
Specifically, we used three broad classes of models with different spacetime metrics. The first class is based on the Kerr solution to Einstein's field equations in GR without charge
\citep{Kerr_1963}. The second class is based on the Kerr-Newman solution to GR, which describes a rotating, charged black hole \citep{Kerr-Newman_metric}.
The third class is a dilaton black hole, which is a solution of the Einstein-Maxwell-dilaton-axion field equations \citep{dilatonBH}. Beyond GR, the dilaton scalar field arises in the low-energy regime of heterotic string theory \citep{PhysRevLett.54.502}.

The forward modeling for the synthetic data generation follows the \ac{eht}-\ac{vlbi} data path to make accurate predictions of what the instrument would measure given a known ground-truth sky brightness distribution.
Multiple realizations of the synthetic datasets are produced to account for uncertain nuisance effects, such as telescope gain errors.
The characteristics of the mock data should be as similar to the observational data as possible.
In \citet{zingularity2}, the second paper in this series, the synthetic data library is used to train a Bayesian neural network for \ac{grmhd}-\ac{grrt} parameter inference and present alongside a demonstration of how this inference can be improved with planned upgrades to the \ac{eht}.
In the third paper \citep{zingularity3}, parameter posteriors are found by applying the trained neural network to current \ac{eht} observations.

This paper is organized as follows:
in \Cref{sec:calibupdate}, we report on improvements made to our \ac{eht} data calibration methods, which are also used for synthetic data generation.
In \Cref{sec:simmodels}, we describe the GRMHD-GRRT image library that contains the ground-truth models of \ac{sgra} and \ac{m87} for our synthetic data generation of \ac{eht} observations.
In \Cref{sec:synthdatagen}, we list all effects taken into account for our forward modeling.
In \Cref{sec:metadata}, we depict the data generation pipeline from an algorithmic point of view together with the content, format, and availability of the produced synthetic data library.
In \Cref{sec:synthdatafeatures}, we describe the salient features of our synthetic data, focusing on characteristics that allow us to discriminate between ground truth model parameters in the presence of the corruption effects that are present in EHT observations.
We offer our conclusions about the GRMHD-GRRT synthetic data library for the EHT in \Cref{sec:synthdataconclusions}.

\section{Updated observational EHT data calibration}
\label{sec:calibupdate}

In this section, we present updates to our observational data reduction with respect to the \ac{m87} and \ac{sgra} calibration methods described in \citet{eht-paperIII} and \citet{eht-SgrAii}, respectively.
For the forward modeling synthetic data generation, we also emulate the \ac{vlbi} data calibration process. This process is based on the updated procedures described below.
Furthermore, the parameter inference presented in \citet{zingularity3} is based on observational data reduced with these updated methods.

\begin{figure}
    \centering
    \includegraphics[width=\columnwidth]{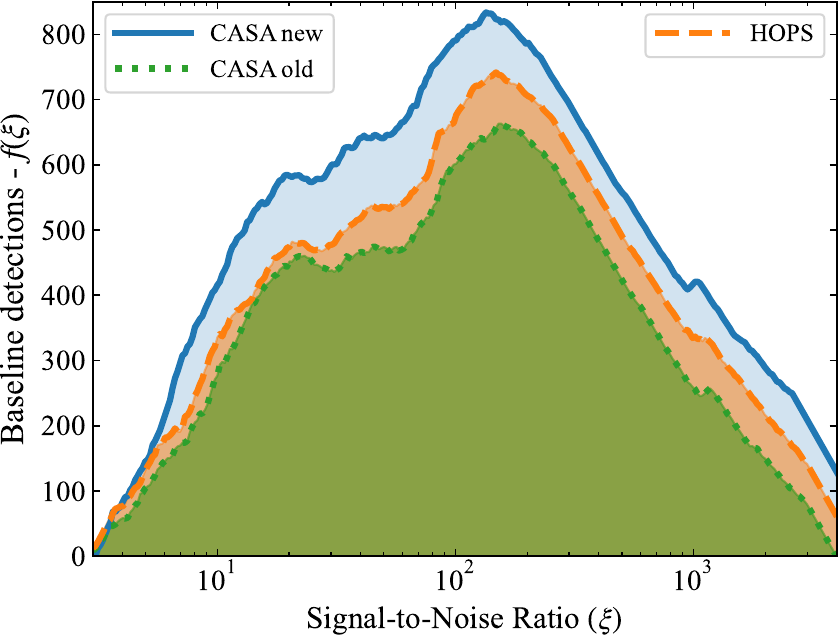}
    \caption{Comparison of the accumulative number of detections of the 226.1\,-\,228.1\,GHz \ac{eht} data of \sgra from 7 April 2017 and \m87 data from 11 April 2017 from different data reductions. The visibilities are averaged into a single frequency channel and time-averaged into 120\,s bins. The signal-to-noise ratio ($\xi$) is computed from the total intensity data (averaged parallel-hand correlation products after the polarization calibration). Only data with $\xi > 3$ are considered as detections, which are counted on all baselines. The detections are plotted as cumulative distributions minus the function $f(\xi) = 280 \log{(\xi)} - 305$.}
    \label{fig:calib_improvement}
\end{figure}

\subsection{Signal stabilization}

Recent upgrades to the \ac{casa} suite \citep{2022CASA,2022Bemmel} and enhancements to the \ac{rpicard} calibration pathway \citep{2019Janssenproc, Janssen2019} have increased fringe detection sensitivity by 10\% across all baselines, mainly at low \ac{sn}, as shown in \Cref{fig:calib_improvement}.
Previously, the signal stabilization through fringe-fitting and atmospheric phase calibration \citep{2022Janssen} was done individually for the parallel-hand correlation products and each of the EHT's 1875\,MHz frequency bands.
Now, after the instrumental phase and delay alignment steps, all four correlation products are combined, each over their full bandwidth, which is 3750\,MHz in aggregate in 2017 and 7500\,MHz in aggregate for the 2018 and subsequent \ac{eht} observations with the current instrumentation.

\subsection{A priori flux density calibration}
\label{sec:apcal}

We fit the telescope gains and estimated their uncertainties for each polarization channel individually, as described in \citet{Janssen2019b} and \citet{eht-paperIII, eht-SgrAii}.
Assuming the primary calibrator sources to have negligible intrinsic circular polarization, the dominant uncertainties from planet models and scatter in the antenna temperature ($T_\mathrm{a}^*$) measurements (due to weather for example) are the same for both polarizations.
From the relative $T_\mathrm{a}^*$ scatter between the polarizations, we estimate that the relative uncertainty is typically at the $1\,\%$ level, which is in agreement with the gain modeling presented in \citet{eht-m87-paper-vii}.
This relative uncertainty is used when drawing gain errors for synthetic data realizations to avoid an overestimation of the data corruption through telescope gains, which could otherwise induce an artificial circular polarization signal in the visibility amplitudes. 
Static, polarization-independent gain offsets are typically at the $\sim$\,10\,\% level \citep{Janssen2019b}.

Moreover, we apply the system temperature measurements prior to the fringe fitting. Thereby, we took measured differences in the sensitivities within signal stabilization solution intervals into account. Additionally, we correctly applied frequency-resolved system temperatures to the corresponding frequency-resolved data. Previously, we have used custom post-processing scripts to apply the flux density calibration metadata to the per-band frequency-averaged data after the signal stabilization. In the case of the \ac{aa}, which provides frequency-resolved calibration measurements at a 58\,MHz level \citep{Goddi2019QA2}, a geometric mean of the system temperatures had previously been applied to the visibilities.

\subsection{Comparison with previous data products}

The combined effect of the aforementioned improvements is visualized in \Cref{fig:calib_improvement}.
For the \sgra data from 7 April 2017 and \m87 data from 11 April 2017, we compare the data previously produced by \ac{rpicard} and \textsc{eht-hops} \citep{Blackburn2019} as described in \citet{eht-paperIII} and \citet{eht-SgrAii} with the improved new \ac{rpicard} data described here.
The old \ac{rpicard}:\ac{casa} data and corresponding old \textsc{eht-hops}:\textsc{hops} data were released under the 10.25739/g85n-f134\footnote{\url{https://doi.org/10.25739/g85n-f134}.} and 10.25739/m140-ct59\footnote{\url{https://doi.org/10.25739/m140-ct59}.} digital object identifiers. For \ac{rpicard}, the old data was obtained with v3.13.1 of the pipeline, which is also containerized with the 
\texttt{30e6ca14fb50275013c668285a3b476f9bc85436\_\newline91da63236db34f3a31b5309b18ac159128f28a35}\footnote{\url{https://tinyurl.com/3p2dy5yw}.} tag. The new data was obtained with \ac{rpicard} v7.2.2, containerized under \texttt{646d6a189c01b04cfa10077a46650038d61687d9\_\newline25c42c3c75a8334d1be4f72bc56b4344dc1f068e}\footnote{\url{https://tinyurl.com/2avph5rd}.}.

The total number of fringe detections is determined by the sensitivity of a data reduction pathway. Imperfect corrections for instrumental phase and delay errors over a telescope bandpass or an incorrect estimation of the thermal noise for example could lead to missed or misidentified fringes.
The distribution of the number of detections versus \ac{sn} is indicative of the data quality. Decoherence effects such as residual delay errors and uncorrected atmospheric phase turbulence can shift detections to lower \ac{sn} for the averaged visibilities. 

From \Cref{fig:calib_improvement}, we observe that all data reduction pathways follow roughly the same distribution of detections.
Compared to the old \ac{casa} data, \textsc{eht-hops} has additional detections to Hawai`i telescopes, which were dropped in the global fringe search of old \ac{casa} as the baselines could not be connected to the remaining telescopes of the \ac{eht} array. 
Additionally, the \textsc{hops} data contains additional detections in the 3 -- 4 \ac{sn} range compared to the old \ac{casa} data. These could either arise from a more conservative fringe rejection threshold used in \ac{casa} or from the robust ad hoc phasing algorithm used in \textsc{eht-hops} to correct for atmospheric phase turbulence \citep{Blackburn2019}. This algorithm can establish sufficient phase coherence for additional detections at low \ac{sn}.

Compared to the old \ac{casa} data, in the new data reduction presented here, sufficient \ac{sn} is accumulated over the combined \ac{eht} bands to connect all Hawai`i fringes to the rest of the array. No detections are lost at low \ac{sn} and excess detections over \textsc{hops} and the old \ac{casa} are obtained at an \ac{sn} of around five. For the other two reduction pathways, the sensitivity and atmospheric phase stabilization are not sufficient to obtain these detections when integrating for 120\,s. The new \ac{casa} data has overall the most detections over the full \ac{sn} range of the data.
Further upgrades of our \ac{eht} data processing will be enabled through the \textsc{ngHOPS} \citep{galaxies10060119} and \textsc{ngCASA} \citep{2022Bemmel, 2022CASA} projects.

\section{Simulation models}
\label{sec:simmodels}

We generated synthetic datasets based on \m87 and \sgra \ac{grmhd}-\ac{grrt} models. These datasets match the 2017 \ac{eht} observations of \m87 \citep{eht-paperIII} and \sgra \citep{eht-SgrAii} in terms of \mbox{($u$, $v$)} coverage and \ac{sn}.
Our ``standard'' parameter set is based on the fiducial \textsc{kharma} Kerr black hole \ac{eht} models from the \textsc{patoka} pipeline \citep{2022Wong, 2021Prather} that are used in \citet{eht-paperV,eht-SgrAv} and ray-traced in all Stokes parameters with \textsc{ipole} \citep{ipole1}.
In Sections~\ref{sec:standard-params-start} to \ref{sec:standard-params-end} below, we describe the broad range of parameters that are sampled by our standard models.
Additional ``exotic'' models used for the synthetic data generation are introduced in \Cref{sec:advanced-params}. The underlying simulations are based on alternative black hole solutions beyond the Kerr metric and sample a smaller parameter space compared to the standard models.
We note here the generally good agreement between the well-tested different GRMHD and GRRT codes used in the \ac{eht} collaboration \citep{2019Porth,2020Gold, 2023Prather}.

\subsection{Magnetic field configuration}
\label{sec:standard-params-start}

Depending on the amount of magnetic flux $\Phi_\mathrm{BH}$ crossing the black hole event horizon, different classes of accretion states develop \citep{eht-paperV, eht-SgrAv}. 
For high $\Phi_\mathrm{BH}$, \ac{mad} states form, when the magnetic flux on the horizon exceeds a threshold where the magnetic pressure becomes larger than the disk's ram pressure. During such a magnetic flux eruption, part of the accumulated magnetic energy is dissapated. Due to the large amount of magnetic flux, powerful jets are launched via the Blandford-Znajek mechanism if the black hole is also rotating \citep{1977Blandford}.
For an accretion rate $\dot{M}$, gravitational radius $r_g$, and speed of light $c$, a \ac{mad} accretion state is reached when $\phi_\mathrm{mag} \equiv \Phi_\mathrm{BH} \left(\dot{M} r_g^2 c\right)^{-0.5} \gtrsim 50$ in Gaussian units. For $\phi_\mathrm{mag} \sim 1$, a \ac{sane} state without flux eruptions develops.

The \ac{grmhd} simulations were initialized with a weakly magnetized Fishbone-Moncrief \citep{1976Fishbone} torus surrounding the black hole.
MAD states were seeded by concentrating the initial magnetic field toward the inner edge of the torus.
The torus is in hydrostatic equilibrium and small perturbations were added to the magnetic field.
Instabilities, such as the magnetorotational instability \citep[MRI,][]{1998Balbus} develop, and the subsequent turbulence triggers accretion onto the black hole. 
Without cooling, this setup self-consistently develops a \ac{riaf} plus jet state.

\subsection{Spin}

For a rotating Kerr black hole with mass $M$ and angular momentum $J$, the dimensionless black hole spin is given as
\begin{equation}
    a_* = \frac{J c}{GM^2}\;,
\end{equation}
where $G$ is the gravitational constant.
Our standard \ac{grmhd} simulations have $a_* = \pm 0.94, \pm 0.5, 0$. The sign of the angular momentum is given with respect to the accretion flow; negative spins correspond to counter-rotation.

\subsection{Inclination angle}

For \sgra, \ac{grrt} images were created for inclination angles of $i_\mathrm{los} = $ \ang{10}, \ang{30}, \ang{50}, \ang{70}, \ang{90}, \ang{110}, \ang{130},\ang{150}, and \ang{170}, with respect to the accretion ﬂow angular momentum vector. Inclinations in the (180, 360) degree range are related to the inclinations in the (0, 180) degree range up to image plane rotation and stochastic effects.

For \m87, the inclination is known to be $\ang{17.2} \pm \ang{3.3}$ based on cm \ac{vlbi} observations \citep{2016Mertens}. Here, we assume that the jet seen on larger scales is aligned with the spin axis of the black hole. In the simulations, $i_\mathrm{los} = \ang{17}$ is used for $a_* < 0$ and $i_\mathrm{los} = \ang{163}$ for $a_* \geq 0$ following the \citet{eht-paperV}.

\subsection{Position angle}

For \sgra, images were rotated by $\theta_\mathrm{PA} = $ \ang{0}, \ang{30}, \ang{60}, \ang{90}, \ang{120}, \ang{150}, and \ang{180} in the plane of the sky. This rotation was performed in the $(u, v)$ plane during the synthetic data generation.

For \m87, the position angle was fixed to \ang{288} based on the orientation of the large-scale jet \citep{Walker2018}. While the large scale jet is known to change orientation on the timescale of years \citep{Walker2018, 2023Cui}, GRMHD model fitting to the 11~April 2017 EHT data of \m87 found $\theta_\mathrm{PA}$ values close to the average \ang{288} \citep{eht-paperVI}, which is also in agreement with close-in-time 3\,mm VLBI observations of the \m87 jet \citep{2023Lu}.

\subsection{Proton-electron temperature coupling}

The single-fluid \ac{grmhd} simulations trace only the temperature $T_i$ of the dynamically important heavier ions.
It is necessary to determine the temperature $T_e$ of the lighter electrons in order to compute the synchrotron emission for the \ac{grrt} images.
Adopted from \citet{2016Moscibrodzka}, the standard models use
\begin{equation}
    T_e = \frac{1+\beta_p^2}{1+\beta_p^2 R_\mathrm{high}} T_i \;.
\end{equation}
Here, $\beta_p$ is the ratio of gas over magnetic pressure.
In the low $\beta_p$ outflow regions, an isothermal jet with $T_e \simeq T_i$ forms (compared to other works, we have fixed $R_\mathrm{low}=1$ here). The free $R_\mathrm{high}$ parameter sets $T_e \simeq T_i/R_\mathrm{high}$ in the accretion disk, where $T_i$ is high due to advective heating and falls off inversely with distance to the black hole.
We set $R_\mathrm{high} = 1, 10, 20, 40, 80, 160$ for \m87 and $R_\mathrm{high} = 1, 10, 40, 160$ for \sgra.

\subsection{Angular scale}

The scale-free \ac{grmhd} simulations were evolved on a numerical grid with a characteristic length given by $r_g$.
For the ray-tracing, the mass $M$ and distance $D$ to the black hole were set to $M_\mathrm{SGRA} = 4.14 \times 10^6\,M_\odot$, $D_\mathrm{SGRA} = 8.127$\,kpc \citep{2019Gravity, 2019Do, 2019Reid} and $M_\mathrm{M87} = 6.2 \times 10^9\,M_\odot$, $D_\mathrm{M87} = 16.9$\,Mpc \citep{2009Blakeslee, 2010Bird, 2011Gebhardt, 2018Cantiello, eht-paperVI, 2022Broderickb, 2023Liepold, 2023Simon} for \sgra and \m87, respectively.
With these parameters, the characteristic angular scales
\begin{equation}
\vartheta = \frac{G M}{c^2 D}\;,
\end{equation}
are $5.0\,\mu$as and $3.6\,\mu$as for \sgra and \m87, respectively.

Following \citet{eht-ml3}, we add $\pm10\,\%$ random variations to $\vartheta$ when generating the synthetic data. These variations reflect uncertainties in $M/D$ and add noise to the data, when they are used for the training of a neural network \citep{zingularity2}. Additional noise occurs in the GRRT models, as physical parameters that would otherwise depend on $M/D$ for a fixed flux density (such as optical depth) are not affected when $\vartheta$ is varied a posetriori \citep[see the discussions in the appendices of][for example]{2021Roelofs, 2022Wong}.

\begin{table*}[t]
\caption{Antenna parameters used for the synthetic data generation.}
\begin{center}
\begin{tabularx}{\linewidth}{@{\extracolsep{\fill}}lccccccccccc}
\hline
\hline
Antenna & $D$ (m) & $\eta_{\textrm{ap}}$ & $\mathcal{S}_\mathrm{rx}$ (Jy) & $\mathcal{G}_\mathrm{err}$ & $\mathcal{D}$ & $\mathcal{P}_\mathrm{rms}$ (") &  $\mathcal{P}_\mathrm{FWHM}$ (") & $\mathcal{W}$ (mm) & $\mathcal{T}_{\mathrm{c}}$ (s) &  $P_{\mathrm{g}}$ (mb) & $T_{\mathrm{g}}$ (K)
\\ \hline
ALMA &  73 & 0.73 & 60   & $1\pm0.03$ & $0\pm0.01$ & 1.0 & 27 & $1.5\pm0.05$ & $10\pm0.5$ & 555 & 271\\
APEX &  12 & 0.63 & 3300 & $1\pm0.05$ & $0\pm0.01$ & 1.0 & 27 & $1.5\pm0.05$ & $10\pm0.5$ & 555 & 271\\
JCMT &  15 & 0.52 & 6500 & $1\pm0.03$ & $0\pm0.01$ & 1.0 & 20 & $1.5\pm0.05$ & $5\pm0.2$ & 626 & 278\\
LMT  &  32 & 0.31 & 2400 & $1\pm0.08$ & $0\pm0.03$ & 1.0 & 10 & $5.0\pm0.5$  & $6\pm0.3$ & 604 & 275\\
PV   &  30 & 0.43 & 1000 & $1\pm0.03$ & $0\pm0.03$ & 0.5 & 11 & $2.3\pm0.1$  & $5\pm0.2$ & 723 & 270\\
SMA  &  16 & 0.73 & 3300 & $1\pm0.03$ & $0\pm0.01$ & 1.5 & 55 & $1.5\pm0.05$ & $5\pm0.2$ & 626 & 278\\
SMT  &  10 & 0.57 & 7700 & $1\pm0.03$ & $0\pm0.03$ & 1.0 & 32 & $4.0\pm0.5$  & $5\pm0.1$ & 695 & 276\\
SPT  &  16 & 0.73 & 3300 & $1\pm0.03$ & $0\pm0.03$ & 1.5 & 55 & $0.8\pm0.05$ & $10\pm0.5$ & 600 & 270\\
AMT  &  15 & 0.52 & 7000 & $1\pm0.05$ & $0\pm0.03$ & 1.0 & 20 & $10\pm0.5$ & $5\pm0.2$ & 772 & 289\\\hline
\end{tabularx}
\label{tab:antparams}
\end{center}
\end{table*}

\subsection{Variability}
\label{sec:standard-params-end}

The \ac{grmhd} fluid models were computed with the $t_\mathrm{g} = G M c^{-3}$ characteristic time cadence, which is about 20\,s for a $4.3\times10^6\,{\rm M}_\odot$ black hole such as \ac{sgra} \citep{2022Gravity} and 8\,h for a $6.5\times10^9\,{\rm M}_\odot$ black hole such as \ac{m87} \citep{2011Gebhardt, eht-paperVI}.
The simulations were evolved until the inner region within about 20 gravitational radii reaches a steady state. In this $t > 10^4\,t_\mathrm{g}$ window, $\sim$1000 frames, spaced with a cadence of $5\,t_\mathrm{g}$, were selected to create \ac{grrt} images for each model. Hereafter, a single model is referred to as a specific set of \ac{grmhd}+\ac{grrt} parameters described above for which we have many individual movie image frames.

We do not expect source variability to occur on timescales smaller than $2\,t_\mathrm{g}$ and hence used single \ac{grrt} images for full 12\,h \ac{eht} observing tracks on \m87. For every \ac{m87} frame of each model, we generated 10 different realizations of synthetic data (see \Cref{sec:synthdatagen} below).
For \ac{sgra}, several consecutive image frames were used for the synthetic data generation of a single few minutes-long \ac{vlbi} scan. To cover a full track, 432 consecutive frames are needed.
From the typically 1000 available images, we can therefore select frames 0 to 568 as possible starting points $f_0$ for a synthetic dataset.
From these, we randomly selected 100 different $f_0$ for each model, each with different realizations of the synthetic data generation parameters.
The mass unit of the accretion flow is a free parameter for the \ac{grrt} images, determined such that the average flux of a \ac{grmhd} run matches the flux measured by the \ac{eht}: 0.5\,\ac{jy} for \m87 \citep{eht-paperIII} and 2.4\,\ac{jy} for \ac{sgra} \citep{eht-SgrAiii,eht-SgrA_LC}.

\subsection{Exotic models}
\label{sec:advanced-params}

\subsubsection{\m87 Kerr-Newman black hole models}

For \m87, we have a set of models based on the Kerr–Newman spacetime metric available. Kerr-Newman spacetimes describe black holes with spin and charge. 
All models have a SANE magnetic field configuration, with different combinations of positive spin values $a_*$ and dimensionless charges $q_*$ in geometrized units:
\begin{itemize}
    \item $a_* = 0$ with $q_*= 0,\; 0.9375$.
    \item $a_* = 0.25$ with $q_*= 0,\; 0.4,\; 0.9$.
    \item $a_* = 0.4687$ with $q_*= 0,\; 0.4,\; 0.8119$.
    \item $a_* = 0.66$ with $q_*= 0,\; 0.33,\; 0.66$.
    \item $a_* = 0.8$ with $q_*= 0,\; 0.2,\; 0.46875$.
    \item $a_* = 0.9375$ with $q_*= 0$.
\end{itemize}
Each of the 14 models was then ray-traced in total intensity with $R_\mathrm{high} = 1, 10, 20, 40, 80, 160$. With 198 frames per model, we have 16,632 images, and for each, multiple \m87 synthetic data realizations were created.
These models will be described in detail in a future publication (Wondrak et al., in prep.).

\subsubsection{\sgra dilaton black hole models}

Typically, non-GR studies pertinent to the \ac{eht} are based on semi-analytical accretion flow models \citep[e.g.,][]{eht-SgrAvi}. So far, only a small number of full GRMHD simulations of non-Kerr spacetimes exist \citep[][]{Mizuno2018,Olivares2020,Roder2022,Roder2022IP}. In this work, we made use of nonrotating dilaton black hole models of \sgra as a representative non-GR spacetime.

The dilaton black hole is described by Einstein-Maxwell-Dilaton-Axion (EMDA) gravity, a class of solutions of low-energy effective string theory \citep{Garcia1995}. The resulting spacetime metric is similar to a Schwarzschild metric, except that it is deformed by the dilaton parameter $b_*$. For more details on EMDA gravity in general, we refer to \citet{WeiLiu2013}, \citet{Flathmann2015}, and \citet{Banerjee2021a,Banerjee2021b}.

In this work, the dilaton parameter was fixed at $b_*=0.504$. That way, the dilaton black hole has the same equatorial innermost stable circular orbit as a Kerr black hole with spin $a_*=0.6$. This value for $b_*$ is consistent with constraints obtained by \citet{eht-SgrAvi} and \citet{2021Kocherlakota}. Two GRMHD simulations with different initial magnetic field configurations, to reach  different accretion states, were carried out using the \textsc{bhac} code \citep{Porth2017,Mizuno2018,Roder2022IP}. 
Next to SANE states with $\phi_\mathrm{mag} \sim 1$, accretion flows develop, where $\phi_\mathrm{mag}$ comes close to $\phi_\mathrm{mag} = 10$. The true MAD state is not reached with the initial setup of these simulations. 

We generated ray-traced images using the \textsc{bhoss} code \citep{Younsi2020a} with $R_\mathrm{high} = 10, 40, 160$ and $i_\mathrm{los} = \ang{20},\ang{40},\ang{60}$. All dilaton images were calculated from purely thermal electron distribution functions in total intensity. From each of the 24 resultant models, 400 Stokes~$\mathcal{I}$ frames were generated with a cadence of $10\,t_\mathrm{g}$. We used groups of 216 consecutive frames, rotated by $\theta_\mathrm{PA} = $ \ang{0}, \ang{30}, \ang{60}, \ang{90}, \ang{120}, \ang{150}, and \ang{180} to generate multiple \sgra synthetic data realizations.

\begin{table*}[t]
\caption{Synthetic data parameter space.}
\begin{center}
\begin{tabularx}{\linewidth}{@{\extracolsep{\fill}}l|ccccc}
\hline
\hline
\hline 
 Category  & Parameter  & \multicolumn{2}{c}{Range of values} &  \multicolumn{2}{c}{Parameter space}   \\ \cline{3-4} \cline{5-6}
  &  & \m87 & \sgra & \m87 &  \sgra  \\
\hline
\hline 
Standard  & $\phi_\mathrm{mag}$ & \textbf{SANE--MAD} &  \textbf{SANE--MAD} &  \textbf{2} &  \textbf{2}  \\
GRMHD & $a_*$ &  $\pm$\textbf{0.94}, $\pm$\textbf{0.5}, \textbf{0} & $\pm$\textbf{0.94}, $\pm$\textbf{0.5}, \textbf{0} & \textbf{10} & \textbf{10}  \\
-\ac{grrt}       & $R_\mathrm{high}$ & \textbf{1}, \textbf{10}, \textbf{20}, \textbf{40}, \textbf{80}, \textbf{160} & \textbf{1}, \textbf{10}, \textbf{40}, \textbf{160} & \textbf{60} & \textbf{40}  \\
models       & $i_\mathrm{los}$ & 17$^\circ$ (163$^\circ$ if $a_* \geq 0)$ & \textbf{10}, \textbf{30}, \textbf{50}, \textbf{...}, \textbf{170}$^\circ$ & 60 & \textbf{360}  \\
       & $\theta_\mathrm{PA}$ & 288$^\circ$ & \textbf{0}, \textbf{30}, \textbf{60}, \textbf{...}, \textbf{180}$^\circ$ & 60 & \textbf{2,520} \\
       & $f_0$ & $0$\,--\,1000 & $0$\,--\,568 & \textit{600,000} & \textit{252,000} \\
\hline
Interstellar & $D_\mathrm{sc}$ & - & $2.7 \pm 0.3$\,kpc & - & \textit{252,000}  \\
scattering   & $R_\mathrm{sc}$ & - & $5.4 \pm 0.3$\,kpc & - & \textit{252,000}  \\
screen       & $\theta_\mathrm{sc,maj}$ & - & $1.380 \pm 0.013 $\,mas & - & \textit{252,000}  \\
toward      & $\theta_\mathrm{sc,min}$ & - & $0.703 \pm 0.013 $\,mas & - & \textit{252,000}  \\
\sgra        & $\phi_\mathrm{sc,PA}$ & - & $\ang{81.9} \pm \ang{0.2}$ & - & \textit{252,000}  \\
             & $\alpha_\mathrm{sc}$ & - & $1.38 \pm 0.06$ & - & \textit{252,000}  \\
             & $r_\mathrm{sc,in}$ & - & $800 \pm 200$\,km & - & \textit{252,000}  \\
\hline
Data         & $\mathcal{P}_\mathrm{rms}$ & \multicolumn{2}{c}{\ang{;;0.5}\,--\,\ang{;;1.5}} & \textit{600,000} & \textit{252,000}  \\
corruption      & $\mathcal{W}$ & \multicolumn{2}{c}{(0.8\,--\,4.0)\,$\pm$\,(0.05\,--\,0.5)\,mm} & \textit{600,000} & \textit{252,000}  \\
effects             & $\mathcal{T}_\mathrm{c}$ & \multicolumn{2}{c}{(5\,--\,10)\,$\pm$\,(0.1\,--\,0.5)\,s} & \textit{600,000} & \textit{252,000}  \\
             & $D_\phi$ & \multicolumn{2}{c}{$\propto \mathcal{T}_\mathrm{c}^{-5/3} $} & \textit{600,000} & \textit{252,000}  \\
             & $\mathcal{D}$ & \multicolumn{2}{c}{1\,--\,3\,\%} & \textit{600,000} & \textit{252,000}  \\
             & $\mathcal{G}_\mathrm{err}$ & \multicolumn{2}{c}{3\,--\,8\,\%} & \textit{600,000} & \textit{252,000}  \\
             & $\sigma_\mathrm{th}^2$ & \multicolumn{2}{c}{$\propto \mathrm{SEFD}_1 \times \mathrm{SEFD}_2$} & \textit{600,000} & \textit{252,000}  \\
\hline
\hline 
\multicolumn{4}{r}{Total number of datasets:} & 600,000 & 252,000 \\
\hline
\end{tabularx}
\label{tab:paramspace}
\end{center}

\textbf{Notes.} -- Parameter space values that are drawn together ($f_0$, interstellar scattering, and data corruption effects) are written in italics. Non-nuisance parameters of physical interest are marked in bold. The scattering parameters are taken from Table~3 in \citet{Johnson2018}. The GRMHD-GRRT parameter space is different for the exotic models (see \Cref{sec:advanced-params}).
\end{table*}

\section{Synthetic data generation process}
\label{sec:synthdatagen}

We used the \symba pipeline \citep{2020Roelofs} to compute synthetic visibilities based on \ac{grrt} images. \symba performs the full forward modeling chain to generate synthetic data that matches the real data as closely as possible \citep[see also][and references therein]{2022Janssen}.
Data corruption effects are added based on first principles with the \meqsilhouette{} software \citep{Blecher2017, Natarajan2021} and the corrupted data are calibrated with the \rpicard{} pipeline \citep{Janssen2019, 2019Janssenproc} in the same way as the real, observational \ac{eht} data.
We summarize the synthetic data generation parameters for every station in the 2017 \ac{eht} array in \Cref{tab:antparams}.
Additionally, we list the Africa Millimeter Telescope \citep[AMT,][]{2016AMT}, with conservative atmospheric parameters, while the exact location for the planned AMT is not yet decided. In this study, we used the location of the 2347m high Gamsberg mountain in Namibia, which is at \ang{23;20;31.9}S \ang{16;13;32.8}E. Results of synthetic data generated with the 2017 \ac{eht} array plus AMT are shown in \Cref{sec:synthdatafeatures}.

The following subsections provide an overview of all effects that are taken into account for the generation of the synthetic dataset, focusing on the overall parameter space of data corruption and calibration processes. More details about the emulated effects, their impact on the data, and justifications for the range of parameters chosen are given in \citet{2020Roelofs}. The \m87 and \sgra data are processed in the same way, unless stated otherwise.

\subsection{Interstellar scattering}

Turbulent magnetized plasma along the line of sight toward the galactic center causes a scattering of radio waves emitted by \ac{sgra} that are received at Earth \citep[e.g.,][]{1976Davies, 1992Langevelde, 2014Bower, Dexter2017}.
The scattering screen results in an angular broadening of the image structure. Additionally, density irregularities in the plasma, that may move across the line of sight, produce variable substructures in the image.

For \ac{sgra} synthetic data, we added interstellar scattering data corruption effects from the \citet{2018Pslatis} and \citet{Johnson2018} model. A random realization with values drawn from the parameters listed in \Cref{tab:paramspace} was created for each observation. More details about the scattering screen toward \ac{sgra} can be found in Section 5.1.2 of \citet{eht-SgrAii}.

\begin{figure}
    \centering
    \includegraphics[width=\columnwidth]{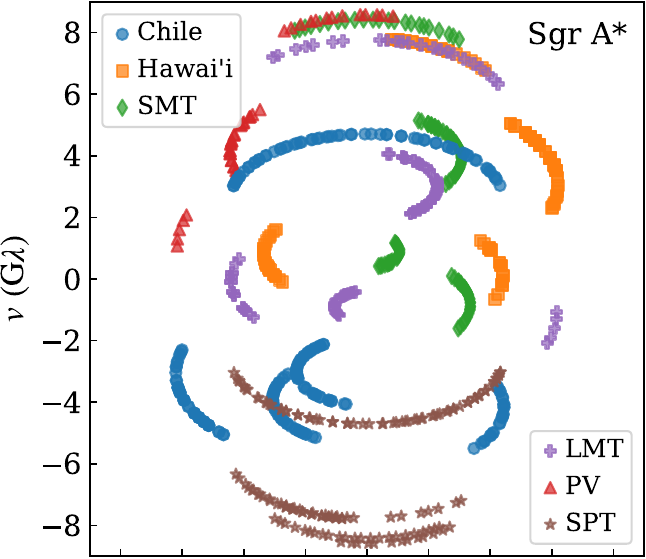}

    \vspace{-0.07cm}
    
    \includegraphics[width=\columnwidth]{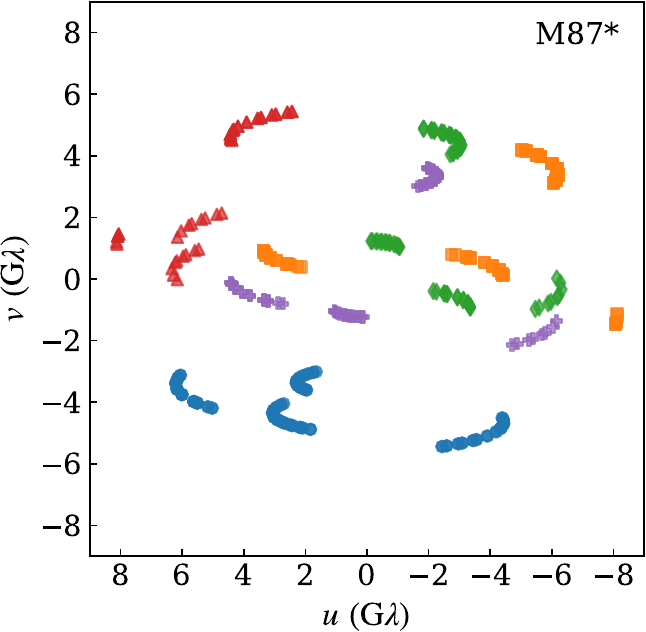}
    \caption{Baseline coverage of the 7 April \sgra (top) and 11 April 2017 \m87 (bottom) 226.1\,--\,228.1\,GHz \ac{eht} data processed with \rpicard{}. The Chile and Hawai'i markers encompass baselines to the co-located ALMA--APEX and JCMT--SMA stations, respectively. The data are averaged over VLBI scan durations and over all frequency channels here and the zero-spacings between co-located sites are not plotted. Conjugate baseline pairs (1-2 and 2-1) are displayed differently following the legends shown in the upper panel.}
    \label{fig:uvcov}
\end{figure}

\subsection{Baseline coverage}

We generated the \m87 and \sgra synthetic data with the $(u, v)$ coverage (\Cref{fig:uvcov}) of the processed 2017 \ac{eht} observations of \m87 on 11 April and \sgra on 7 April \citep{eht-paperIII, eht-SgrAii}, respectively.
Fringe non-detections and other data dropouts were thereby taken into account.

For the \sgra observations, the \ac{aa}, \ac{ap}, \ac{jc}, \ac{lm}, \ac{pv}, \ac{sm}, \ac{az}, and \ac{sp} observatories participated in the observations.
\m87 was observed by the same telescopes, minus the \ac{sp}, which cannot see the source. 

\subsection{Antenna pointing errors}

We used an ad hoc prescription to model antenna pointing errors.
Telescope beam offsets $\rho$ are drawn from a normal distribution centered around zero with a standard deviation of $\mathcal{P}_\mathrm{rms}$. These offsets introduce stochastic variability on a specified atmospheric coherence time $\mathcal{T}_\mathrm{c}$ and gross per-scan amplitude losses, for which $\rho$ slowly increases by $3\,\%$ per scan until new pointing offsets are drawn every $\sim 5$ scans (indicating that telescopes have obtained new pointing solutions for their dishes during the observing run).

We assumed Gaussian telescope beams with a full-width at half maximum of $\mathcal{P}_\mathrm{FWHM}$, for which fractional amplitude losses $\Delta z$ are given by
\begin{equation}
    \Delta z =  \mathrm{exp}\left(-4\ln{2}\frac{\rho^2}{\mathcal{P}_{\mathrm{FWHM}}^2}\right)\, .
\end{equation}

\subsection{Earth atmosphere}
\label{sec:earthatm}

We use the atmospheric module of \meqsilhouette{} to simulate four data corruption effects:
\begin{enumerate}
    \item The Atmospheric Transmission at Microwaves (\atm{}) software \citep{Pardo2001} is used to solve the radiative transfer equation and to compute the amount of signal attenuation along the line of sight through the Earth's atmosphere. \atm{} is initialized based on the amount of precipitable water vapour $\mathcal{W}$ at zenith, the local pressure $P_\mathrm{g}$, and temperature $T_\mathrm{g}$ at each antenna. The values for these quantities were taken from EHT observations logged in the $\textsc{vlbimonitor}$ \citep{eht-paperII}.
    \item \atm{} also computes the amount of sky noise $T_\mathrm{sky}$ arising from the radiation produced by the atmosphere at the observing frequency.
    \item Wet dispersive path delays are simulated with \atm{}.
    \item As described in Section~3.2.2 of \citet{Natarajan2021}, Kolmogorov phase turbulence $\delta \phi$ from the troposphere is simulated with a power law approximation of the phase structure function $D_\phi \propto \mathcal{T}_\mathrm{c}^{-5/3}$ for a zenith atmospheric coherence time $\mathcal{T}_\mathrm{c}$. The time series of phase errors are $\propto \mu \nu \, \delta \phi$, given the airmass $\mu$ toward the horizon and observing frequency $\nu$. We simulated visibilities with 32 channels, spanning a bandwidth of 2\,GHz around a central frequency of 227\,GHz, where the ray-tracing was performed to make the \ac{grrt} images. Compared to previous simulation data from \meqsilhouette{}, where $\delta \phi$ itself was also approximated with a power-law \citep{Blecher2017}, we now use a Cholesky factorization in \meqsilhouette{} v2 \citep{Natarajan2021}.
\end{enumerate}

\subsection{Thermal noise}

Thermal noise is determined by the System Equivalent Flux Densities (SEFDs) of every antenna, which measure the total noise contribution along the signal path.
We computed the SEFDs as
\begin{equation}
    \mathrm{SEFD} = \mathcal{S}_\mathrm{rx} + 8 \frac{k_\mathrm{B} T_\mathrm{sky}}{\pi \eta_\mathrm{ap}D^2}\,,
    \label{eq:sefds}
\end{equation}
with $\mathcal{S}_\mathrm{rx}$ the noise contribution from the telescope's receiver, $k_\mathrm{B}$ the Boltzmann constant, $D$ the diameter of the telescope dish, and $\eta_\mathrm{ap}$ the aperture efficiency.

Noise was added to the visibilities by randomly drawing from a Gaussian probability distribution function with a standard deviation of
\begin{equation}
\sigma_\mathrm{th}=\frac{1}{\eta_\mathrm{Q}}\sqrt{\frac{\mathrm{SEFD}_1{\mathrm{SEFD}_2}}{2\,\mathrm{GHz} \times 0.5\,\mathrm{s}}}\,.
\end{equation}
Here, $\eta_Q = 0.85457$ is the \ac{eht} quantization efficiency for the CASA-based \ac{eht} data reduction.

\subsection{Polarization leakage}

We modeled residual polarization leakage ($\mathcal{D}$-term) effects based on the accuracy at which we can constrain their magnitude for the \ac{eht} \citep{eht-m87-paper-vii}.
We assume the $\mathcal{D}$-terms to be constant over observing tracks as well as the frequency bandwidth and add noise by randomly drawing from a Gaussian probability distribution function with standard deviation $\sigma_\mathcal{D}$ for each station.
For the co-located \ac{aa}, \ac{ap} and \ac{jc}, \ac{sm} sites, we used $\sigma_\mathcal{D} = 0.01$. For all other antennas, where the leakages are more difficult to constrain \citep[see, e.g.,][]{2022Issaoun,2023Jorstad}, we used $\sigma_\mathcal{D} = 0.03$.

\subsection{Telescope amplitude gain errors}

In addition to antenna pointing errors, we simulated gross gain errors $\mathcal{G}_\mathrm{err}$ for each station.
These arise mostly from gain measurement errors in the real data.
Typical gain errors are $\mathcal{O}(3\,\%)$ at a 1-sigma level \citep{Janssen2019b}. The \ac{lm} has the largest gain uncertainties of $\sim 8\,\%$.
The gains are mostly dependent between the RCP and LCP circular polarization feed signal paths of the telescopes. We based LCP gain offsets on a random draws of RCP offsets with an additional $\mathcal{O}(1\,\%)$ random relative offset (\Cref{sec:apcal}).
Examples of residual self-calibration gains from the imaging of \ac{eht} data are given in Table~14 of \citet{eht-paperIV}.

\subsection{Phase calibration}

The path delay and phase turbulence corruption effects added by the simulated Earth atmosphere (\Cref{sec:earthatm}) have to be corrected to obtain a realistic dataset.
The total flux density and source structure of the input model image determine the correlated flux density across the interferometric baselines and thereby how well these effects can be corrected in the forward modeling.
Fringe-fitting \citep{2017Thompson} the data to solve for delay offsets determines if the source is detected for particular scans and antennas.
Intra-scan phase turbulence is corrected by fringe-fitting segmented pieces of data.
The length of the segmentation intervals is determined by the \ac{sn} on baselines to a chosen reference station \citep{Janssen2019}. Akin to the real data, this left residual atmospheric phase wander in low \ac{sn} data (on long baselines or near interferometric nulls), which causes decoherence effects when the data are averaged in time.

\subsection{Amplitude calibration}

Accurate a priori models for the flux densities of resolved and variable \ac{vlbi} sources are typically not available.
Commonly, the amplitude calibration is done with a priori estimations for each station's total noise budget instead (Equation~\ref{eq:sefds}).
For continuum high-frequency \ac{vlbi} observations, the sky temperature is the only substantial variable source of noise.
In good weather conditions, $T_\mathrm{sky}$ only varies with the amount of arimass toward the horizon, which is set by the telescope elevation angle.
Usually, single load chopper calibration measurements are performed by every station right before any \ac{vlbi} scan to measure the total noise and atmospheric attenuation.
We emulated these measurements by correcting the \atm{} attenuation $\tau$ of every scan with the value of $\tau$ given at the start of the scan, leaving intra-scan variations of the attenuation due to small changes in elevation angle uncorrected.
Higher-order noise contributions, which could arise due to spillover at the telescope or the contribution from the astronomical source itself, for example, are ignored here.

Additionally, the network-calibration technique \citep{Blackburn2019} is used in the real data to accurately constrain the gains of co-located sites based on
\ac{aa} and \ac{sm} measurements of the total source flux density at large scales.
The same flux is assumed to be measured on the short \ac{aa}-\ac{ap} and \ac{jc}-\ac{sm} \ac{eht} baselines, both of which do not resolve structures smaller than $\sim$100\,mas.
For the time-variable \sgra models, a time-dependent light curve \citep{eht-SgrA_LC, eht-SgrAii} is constructed and used to constrain gains at the time cadence of the individual input model movie frames, mimicking the network-calibration employed for the  observational \sgra data.
For the synthetic data, we assumed that \ac{aa} calibration errors are negligible for the network calibration, that is, we used the uncorrupted model fluxes.

\section{Data workflow, content, format, and availability}
\label{sec:metadata}

\begin{figure*}[t]
    \centering\offinterlineskip
    \includegraphics[height=14.6cm]{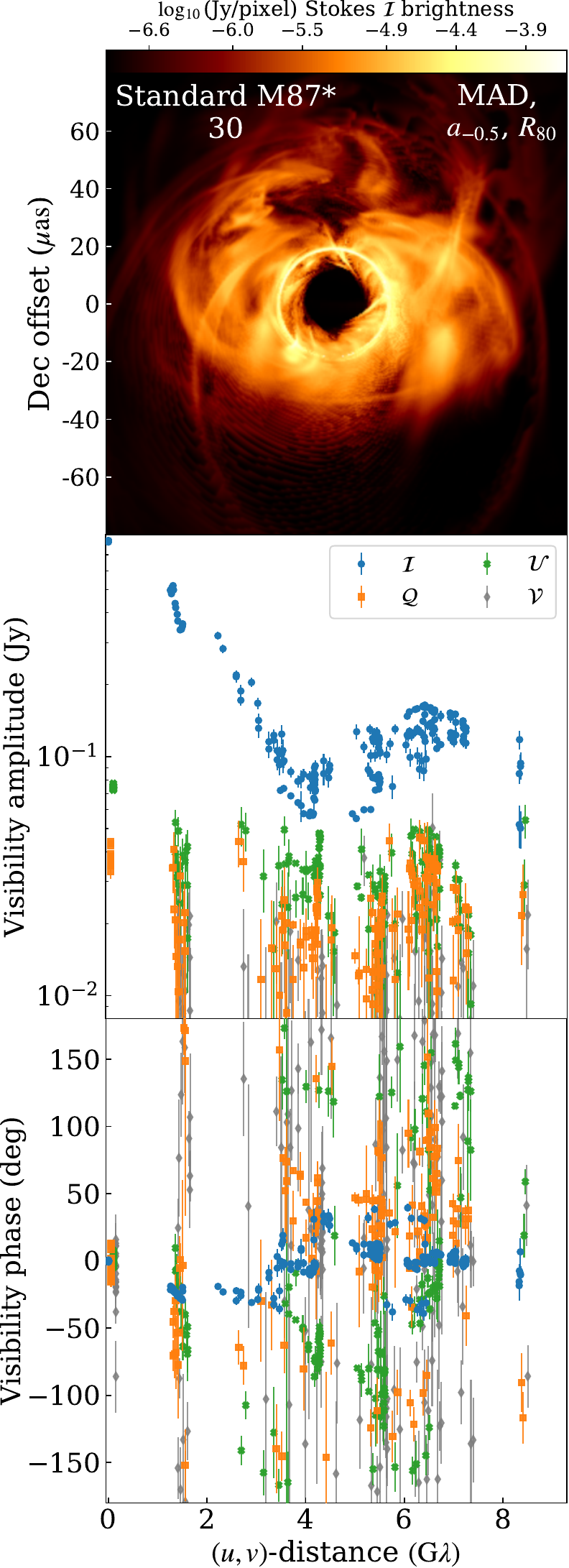}
    \hskip -0.6ex\includegraphics[height=14.6cm]{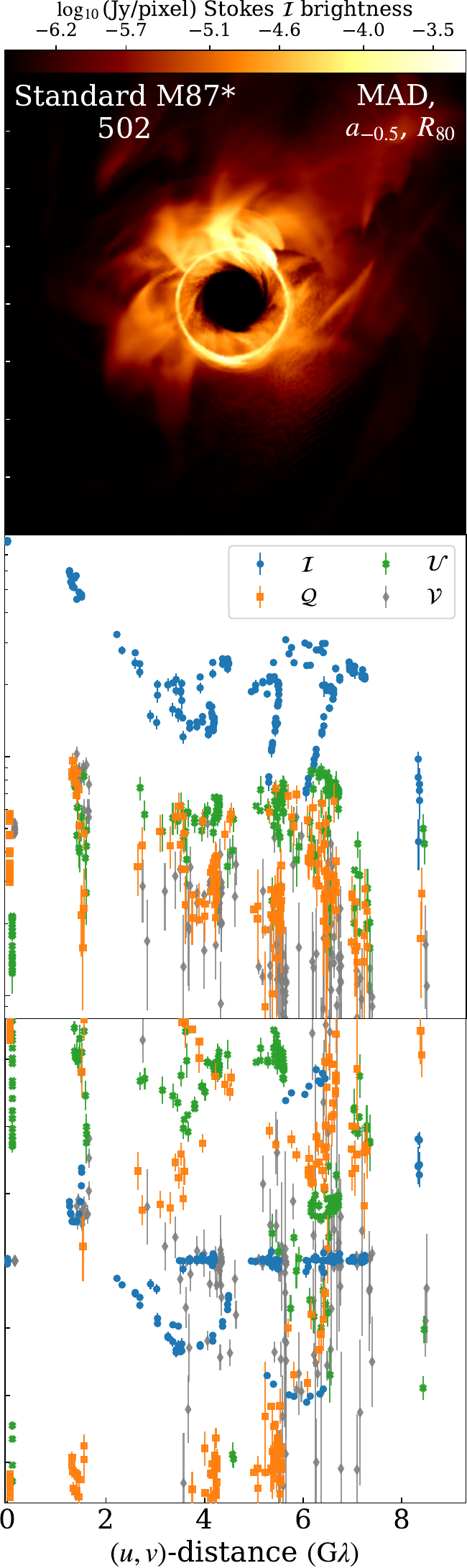}
    \hskip -0.6ex\includegraphics[height=14.6cm]{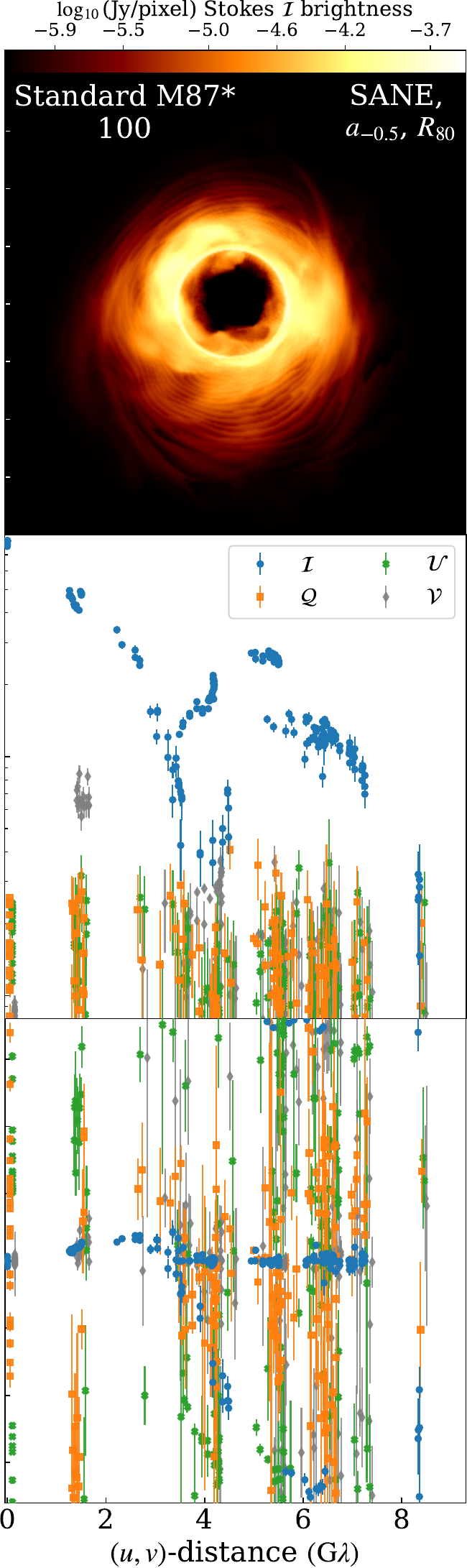}
    \hskip -0.6ex\includegraphics[height=14.6cm]{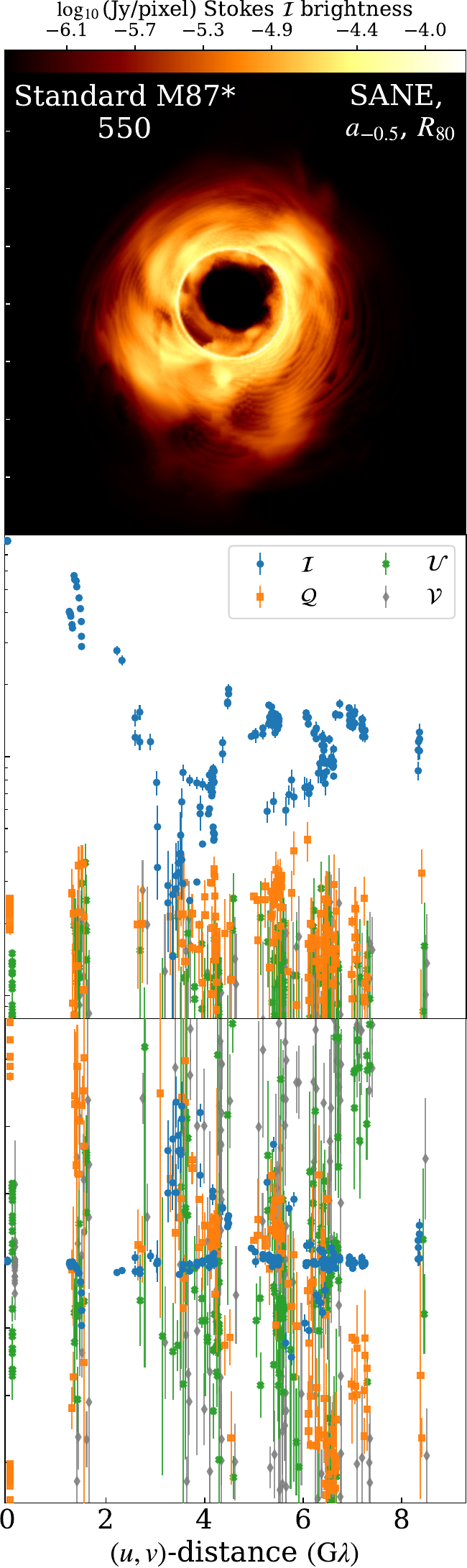}
    \caption{Four synthetic datasets based on two realizations of two standard \m87 models are presented. GRRT frame numbers are displayed in the top left corners. The top row shows the total intensity ray-traced ground-truth model images on logarithmic scales with varying dynamic ranges. Visibility amplitudes on a logarithmic scale and phases of corresponding synthetic data realizations are displayed with thermal noise error bars as a function of baseline length in the middle and bottom rows, respectively. The measurements shown can come from different orientations at the same baseline length. For better readability, the visibilities have been averaged over scan durations, amplitudes lower than 0.008\,\ac{jy} have been clipped, and the values of the different Stokes parameters are each offset by $50\,\mathrm{M}\lambda$ on the x-axis. Spin $a_*=s$ and $R_\mathrm{high}=r$ parameters are listed in a shorthand notation as $a_s$ and $R_r$ in the top-right corner of each model image.}
    \label{fig:synthetic_data_m87}
\end{figure*}

\begin{figure*}
    \centering\offinterlineskip
    \includegraphics[height=14.6cm]{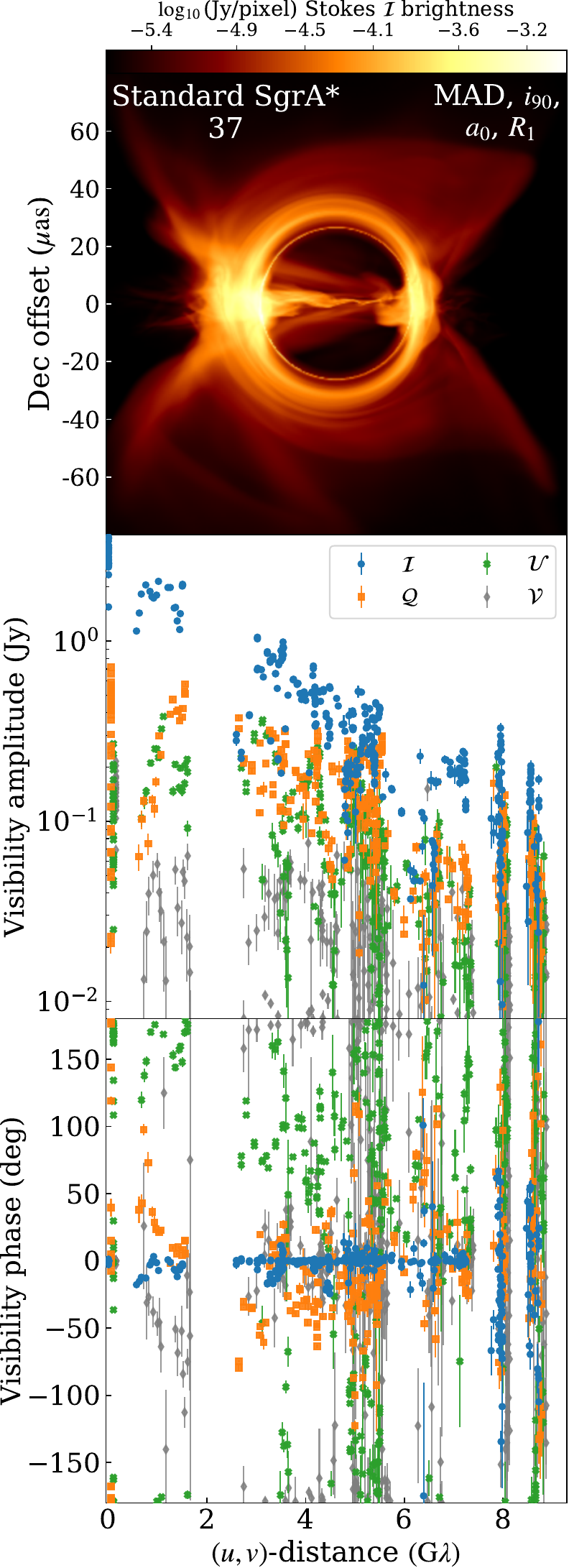}
    \hskip -0.6ex\includegraphics[height=14.6cm]{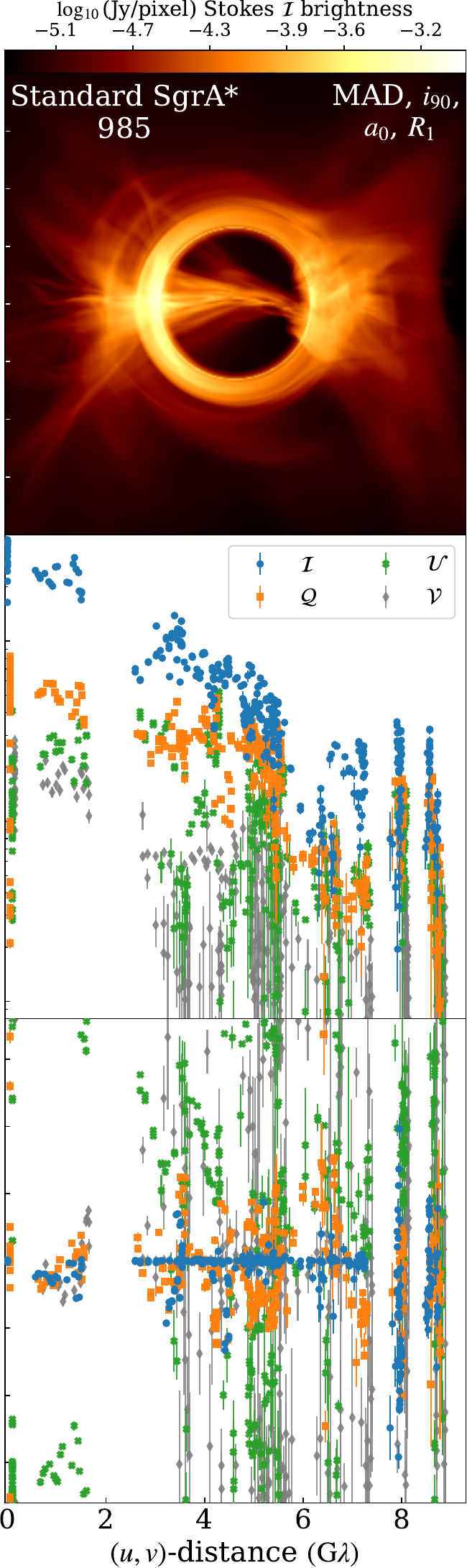}
    \hskip -0.6ex\includegraphics[height=14.6cm]{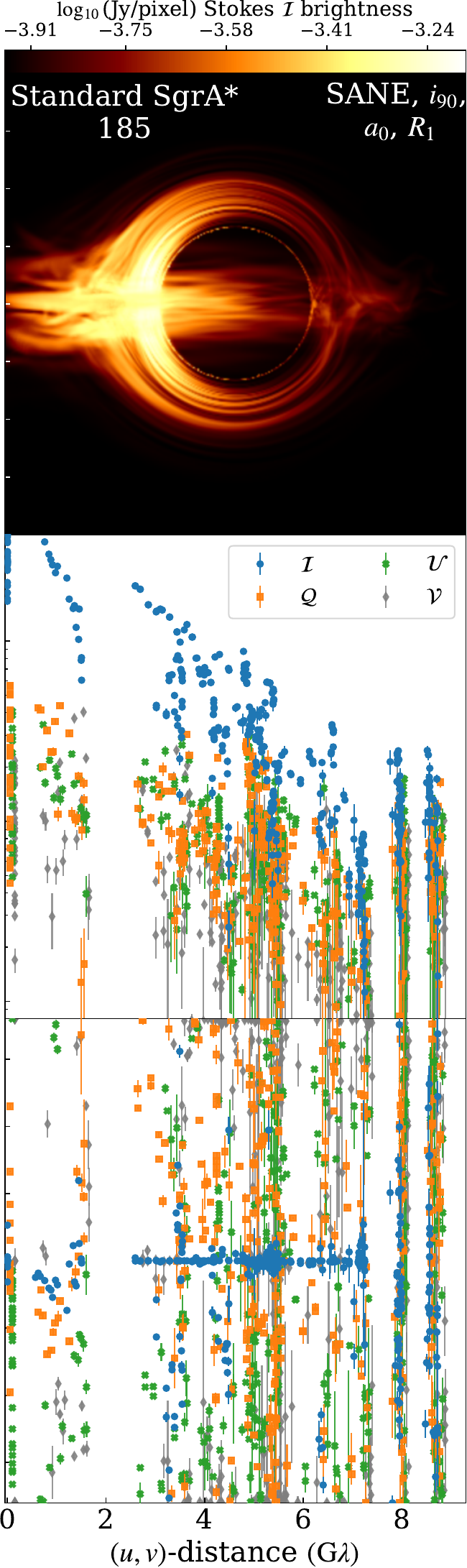}
    \hskip -0.6ex\includegraphics[height=14.6cm]{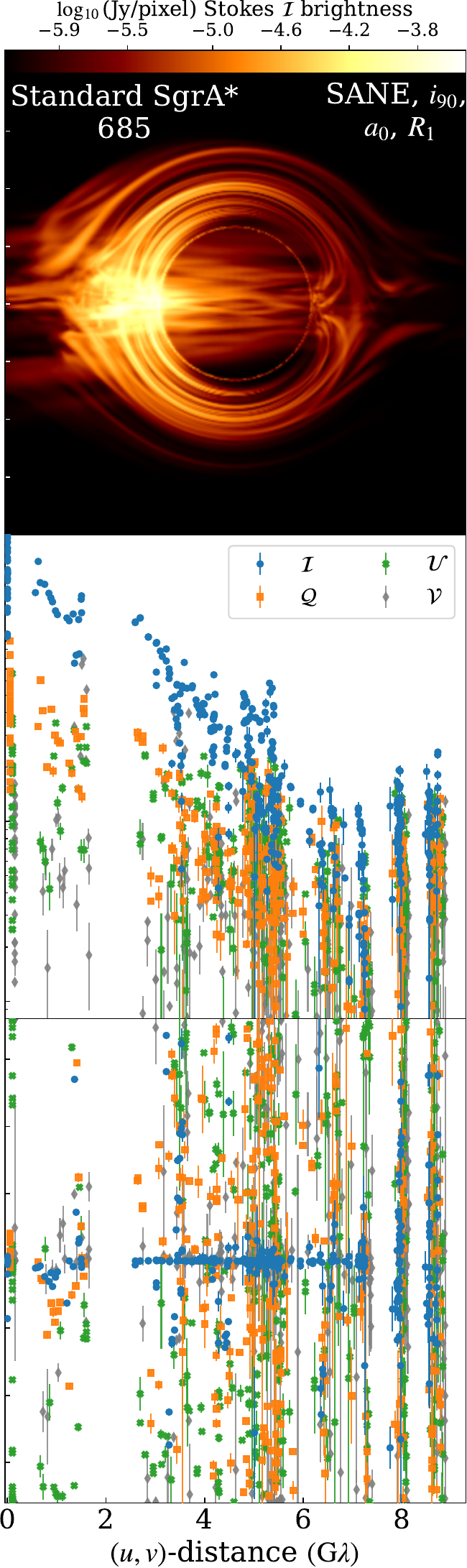}
    \caption{Same as \Cref{fig:synthetic_data_m87} but for $\theta_\mathrm{PA} = 0$ standard \sgra models, where the $i_\mathrm{los}=l$ parameter is indicated with a $i_l$ label. Single representative GRRT frames are shown for the data that are built up from a movie of many consecutive frames. Zero-baseline fluxes above 3.9\,\ac{jy} have been clipped.}
    \label{fig:synthetic_data_sgra}
\end{figure*}

\begin{figure*}
    \centering\offinterlineskip
    \includegraphics[height=14.6cm]{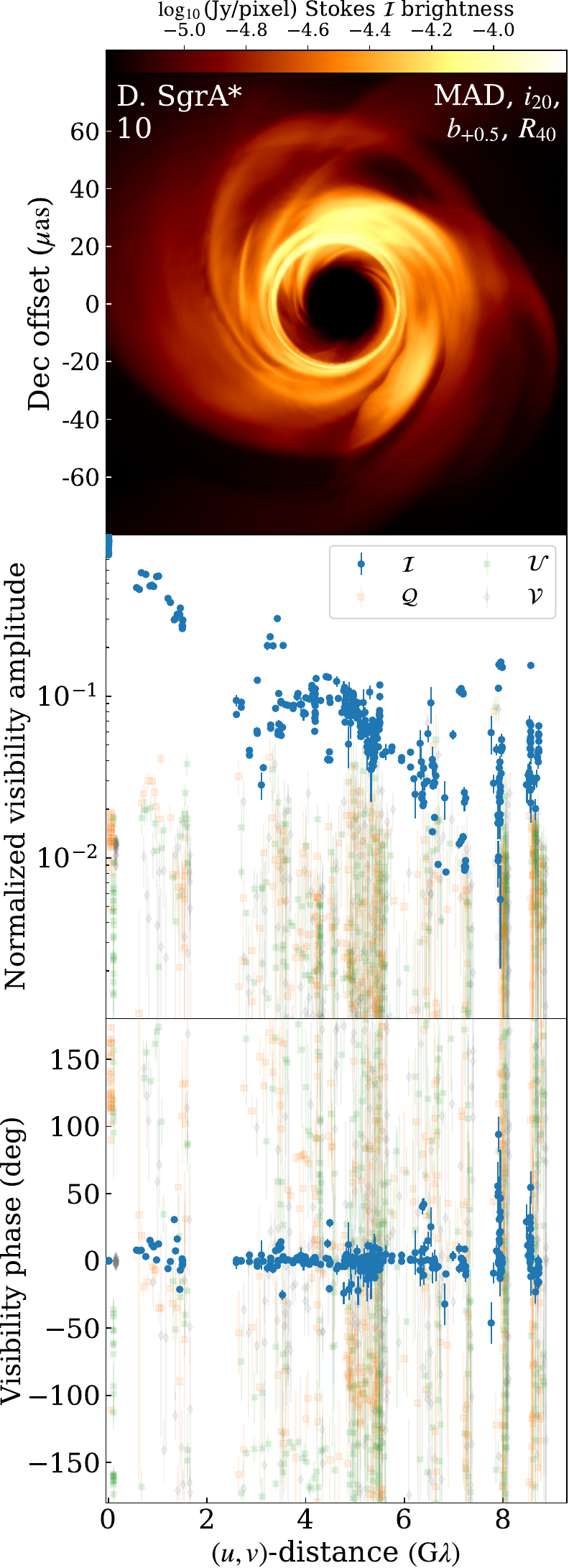}
    \hskip -0.6ex\includegraphics[height=14.6cm]{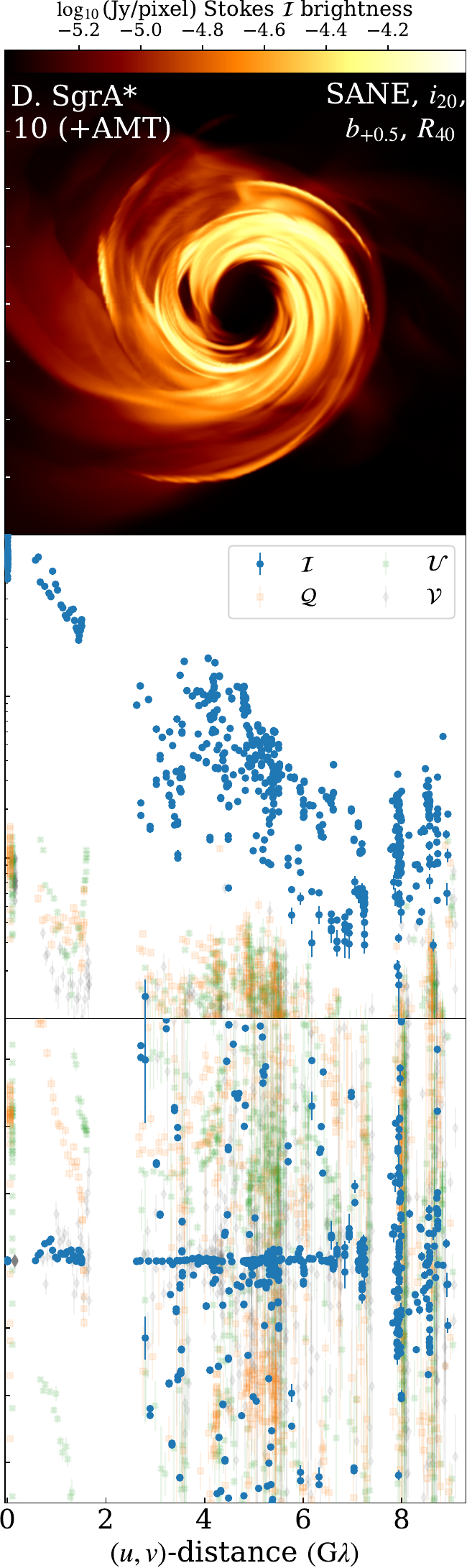}
    \hskip -0.6ex\includegraphics[height=14.6cm]{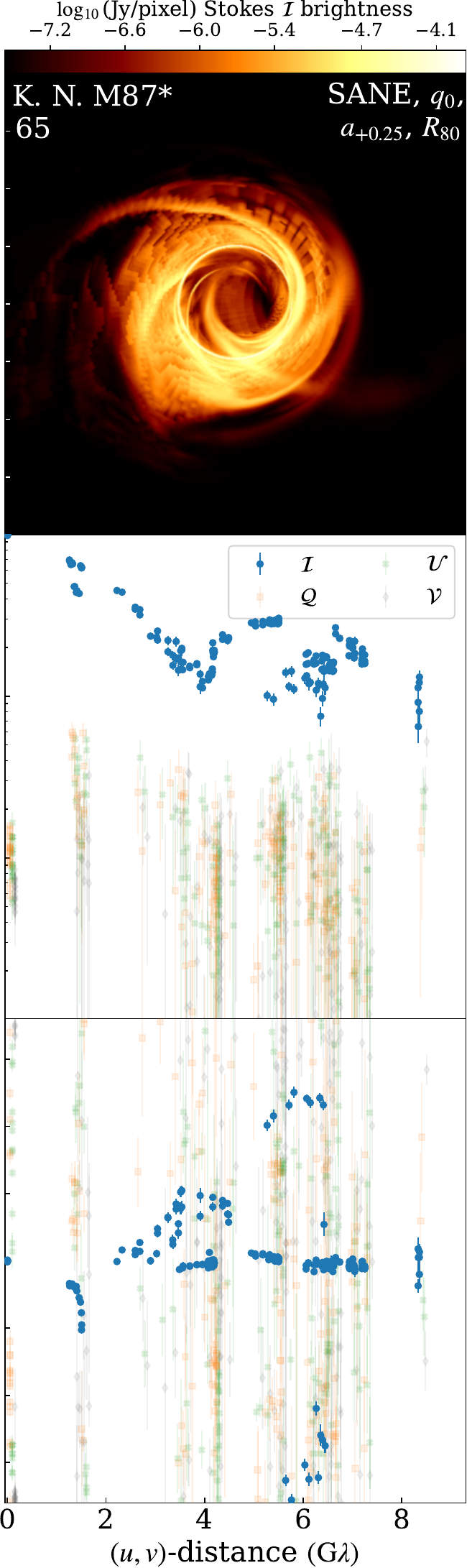}
    \hskip -0.6ex\includegraphics[height=14.6cm]{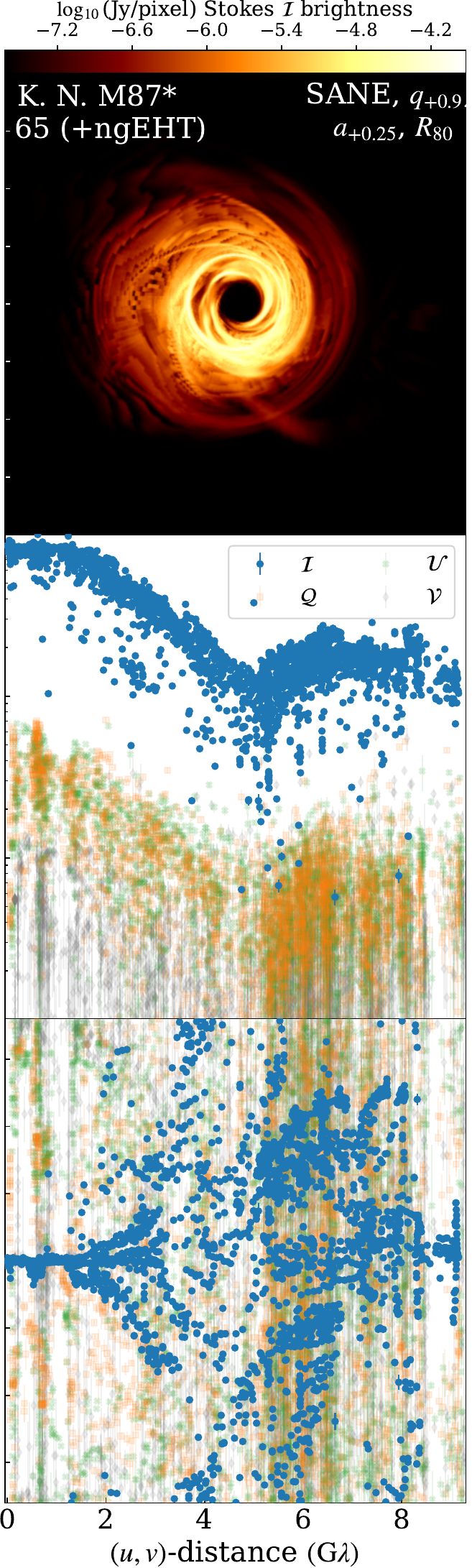}
    \caption{Similar to \Cref{fig:synthetic_data_m87} and \Cref{fig:synthetic_data_sgra} for the \m87 and \sgra models, respectively. Here, we have \sgra dilaton (D.) and \m87 Kerr-Newman (K. N.) models (\Cref{sec:advanced-params}). For the former, the dilaton parameter $b_*=0.504$ (see text) is labeled as $b_{+0.5}$. For the latter, the geometrized charge $q_*=C$ is given with a $q_C$ label. As the D. and K. N. models were ray-traced only in total intensity, we have faded out the Stokes $\mathcal{Q}, \mathcal{U}, \mathcal{V}$ data. Visibility amplitudes have been normalized to unity. Next to the 2017 coverage, measurements are shown for possible future \ac{eht} observations, where the AMT or ngEHT would join (see text).}
    \label{fig:synthetic_data_dkn}
\end{figure*}

The input models and output synthetic datasets are staged on the CyVerse (formerly the iPlant Collaborative) data storage \citep{cyverse1, cyverse2} and the synthetic data generation software \symba is run on the Open Science Grid \citep{osg07, osg09, osgdoi1, osgdoi2} as well as the Cobra HPC system at the Max Planck Computing and Data Facility.
The \textsc{Pegasus} workflow management system \citep{deelman-fgcs-2015} is used to schedule the computations and handle the data exchange between CyVerse and the Open Science Grid in a reproducible manner.
Access to the initial model data and final synthetic data on CyVerse is facilitated through \texttt{iRODS}.\footnote{\url{https://irods.org}.}

We simulated data for all four correlation products. For circular feeds, we have the RR, LL parallel-hand and RL, LR cross-hand complex visibilities measured on all baselines (except for the \ac{jc}, which only observed single-pol data in 2017).
The Stokes parameters can be extracted from these measurements as
\begin{eqnarray}
\label{stokes}
\mathcal{I} & = & \frac{1}{2} \left( \mathrm{RR} + \mathrm{LL} \right) \\
\mathcal{Q} & = &\frac{1}{2} \left( \mathrm{RL} + \mathrm{LR} \right) \\
\mathcal{U} & = &\frac{i}{2} \left( \mathrm{LR} - \mathrm{RL} \right) \\
\mathcal{V} & = &\frac{1}{2} \left( \mathrm{RR} - \mathrm{LL} \right) \; ,
\end{eqnarray}
with $i=\sqrt{-1}$.

\symba creates frequency-resolved synthetic data in the \texttt{MeasurementSetv2}\footnote{\url{https://casa.nrao.edu/Memos/229.html}.} format.
We do not store these datasets permanently; instead, we store the frequency-averaged data in the \texttt{UVFITS}\footnote{The \texttt{UVFITS} format is described in \url{ftp://ftp.aoc.nrao.edu/pub/software/aips/TEXT/PUBL/AIPSMEM117.PS}. It is coupled to a specific version of the AIPS software (\url{http://www.aips.nrao.edu}) and undergoes minor revisions from time to time.} format on CyVerse. We have used the \texttt{{49a813d2dc62eac809f3909bee0d38a8b113ffc4}}\footnote{\url{https://hub.docker.com/r/mjanssen2308/symba}.} \symba \textsc{Docker}\footnote{\url{https://www.docker.com/}} container to generate the synthetic date presented in this work.

As our final data product, we bundled the complex correlation coefficients for groups of 1000 synthetic datasets together, each labeled with $\phi_\mathrm{mag}, a_*, R_\mathrm{high}, i_\mathrm{los},$ and $\theta_\mathrm{PA}$ in single \texttt{TFRecord}\footnote{\url{https://www.tensorflow.org/tutorials/load_data/tfrecord}.} files.
These files efficiently serialize structured data in a coding language- and hardware platform-independent manner based on Google's \texttt{protocol buffer}\footnote{\url{https://developers.google.com/protocol-buffers}.} method.
The self-contained \texttt{TFRecords} do not allow for random data access but take up less disk space, are easier to handle compared to the large number of individual \texttt{UVFITS} files, allow for rapid parallel I/O operations, and consistently store data together with their labels.
Machine learning frameworks such as \tensorflow{} \citep{tf1, tf2} can combine any number of individual \texttt{TFRecord} files into single objects, for which data can be loaded in optimally sized chunks based on the amount of available system memory and operations such as shuffling and batching are easily performed.
The real and imaginary components of the visibilities are stored in one-dimensional arrays sorted by $(u, v)$-distance for each correlation product. In cases where fringe non-detections occur in the forward modeling process, the flagged visibilities are replaced with zeroes.

The single data set \texttt{UVFITS} files have a size of 652\,Kilobytes and 1.5\,Megabytes for \m87 and \sgra, respectively. The corresponding \texttt{TFRecord} file sizes of 1000 bundled datasets are 168 and 423\,Megabytes, respectively.
For the full parameter space shown in \Cref{tab:paramspace}, there are about 100\,Gigabytes of \texttt{TFRecord} data for each source.
Access to the \texttt{UVFITS} and \texttt{TFRecord} files will be given upon reasonable request.
The availability of the observational \ac{eht} data and corresponding processing software is described in \citet{zingularity3}.

\section{Synthetic data features}
\label{sec:synthdatafeatures}

\begin{figure}[t]
    \centering
    \includegraphics[width=\columnwidth]{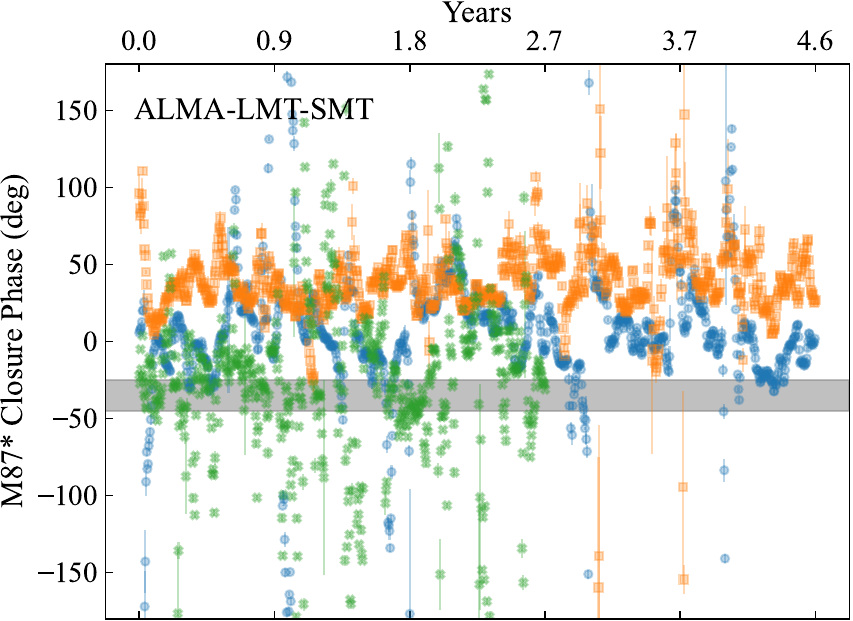}
    \includegraphics[width=\columnwidth]{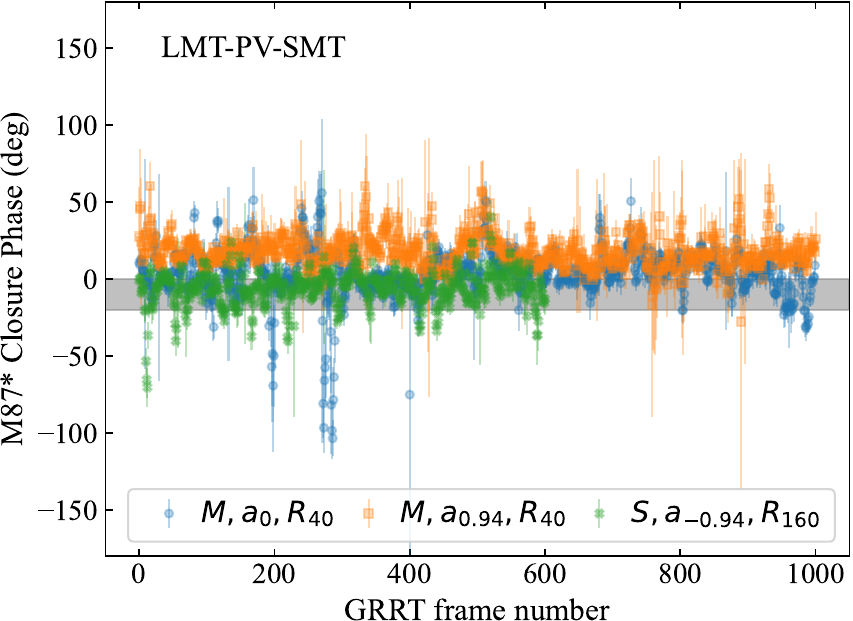}
    \caption{Total intensity closure phase evolution of \m87 synthetic data from example MAD ($M$) and SANE ($S$) standard models as a function of model variability for the ALMA-LMT-SMT and LMT-PV-SMT triangles, respectively. Spin $a_*=s$ and $R_\mathrm{high}=r$ parameters are listed in a shorthand notation as $a_s$ and $R_r$ in the legend of the bottom panel. Each data point corresponds to a VLBI scan-averaged synthetic data closure phase measured at 02:27 UT on 11 April 2017 and the standard deviations plotted are computed from multiple synthetic data realizations of the same model frame. The corresponding measurement range from the 2017 observational data is depicted with a gray band. The SANE model shown here has been ray-traced for 600 frames only.}
    \label{fig:m87_cphase_vs_time}
\end{figure}

\begin{figure}[t]
    \centering
    \includegraphics[width=\columnwidth]{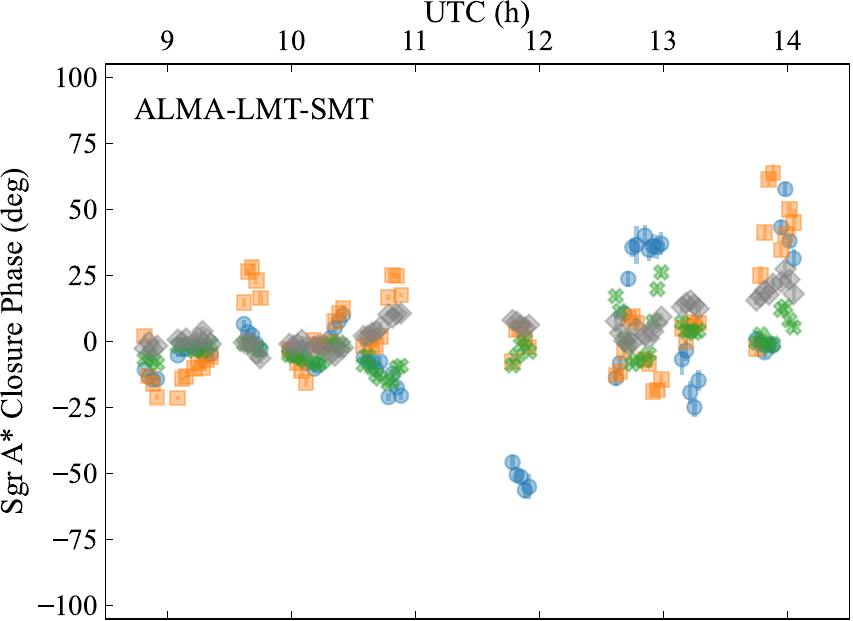}
    \includegraphics[width=\columnwidth]{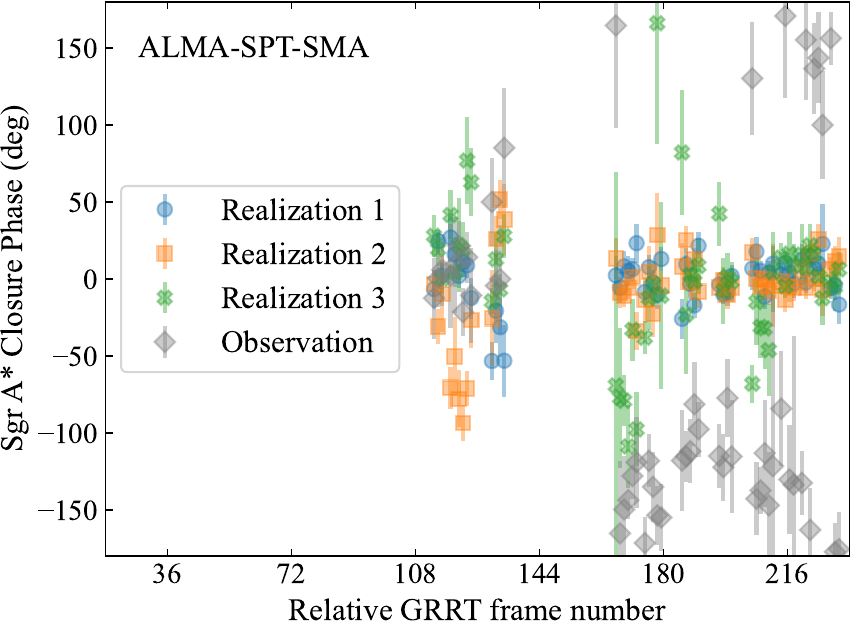}
    \caption{ALMA-LMT-SMT and ALMA-SPT-SMA total intensity closure phase evolution of 120\,s averaged \sgra data over the course of the 2017 \ac{eht} observing track on 7 April. Three synthetic data realizations of the variable MAD, $a_*=0.5$, $R_\mathrm{high}=160$, $i_\mathrm{los}=\ang{30}$, $\theta_\mathrm{PA}=0$ model are shown next to the observational data. The error bars depict the a priori thermal noise estimations of the data.}
    \label{fig:sgra_cphase_reals}
\end{figure}

\Cref{fig:synthetic_data_m87} depicts example synthetic data realizations for four different \m87 GRRT frames with $a_*=-0.5$ and $R_\mathrm{high} = 80$. The two realizations of the more variable MAD model show more extended emission compared to the two SANE realizations.
The synthetic visibilities displayed here are averaged over VLBI scan durations and baseline orientations. 
We compared the differences between the data realizations of the same model with synthetic data, where thermal noise was added as the only data corruption effect.
We found that the differences in the total intensity data are primarily caused by intrinsic model variations, while differences in polarization-sensitive visibilities are mostly due to differences in the simulated data corruption effects, the residual $\mathcal{D}$-terms for example.
The strong circular polarization amplitude at about 1.8\,G$\lambda$ in the SANE GRRT \#\,100 model realization for instance is the result of telescope gain errors.
For the \Cref{fig:synthetic_data_m87} case study, the presence of a prominent total intensity visibility minimum at about 3.5\,G$\lambda$ is a distinctive data feature only present in the SANE model. We show and discuss the differences between the \symba simulated synthetic data and ground truth from the underlying model image in \Cref{app:SvG}.

\Cref{fig:synthetic_data_sgra} presents example synthetic data realizations from the highly variable \sgra models; here two MAD and two SANE realizations with $a_*=0$, $R_\mathrm{high} = 1$, $i_\mathrm{los} = \ang{90}$, and $\theta_\mathrm{PA}=0$.
These models show edge-on disk-dominated synchrotron emission in a Schwarzschild spacetime.
In this case, salient features can be found in the displayed linear polarization visibility phases. Compared to the SANE model, the MAD's $\mathcal{Q}$ and $\mathcal{U}$ phases have a distinct structure and are more coherent with baseline length.

\Cref{fig:synthetic_data_dkn} shows synthetic data for alternative \sgra and \m87 models (\Cref{sec:advanced-params}) for both the 2017 baseline coverage as well as possible future \ac{eht} observations, where the AMT or ngEHT array \citep{2019ngEHT} join the EHT.
For \sgra \ac{eht} observations, the AMT adds substantial improvements to the baseline coverage \citep{2023LaBella}. Currently, the longest baseline on our primary targets is \ac{pv}--\ac{sp} with $8.7\,\mathrm{G}\lambda$ at 230\,GHz. With the AMT, we will have \ac{az}--AMT out to $8.9\,\mathrm{G}\lambda$ at 230\,GHz with an improved northeast and southwest resolution.
To demonstrate the capabilities of a significantly enhanced EHT array, comparable to the 23 stations that can currently observe at 86\,GHz with the Global mm-VLBI Array \citep[GMVA,][]{2006Krichbaum}, we have simulated observations of 22 dishes from the combined current EHT plus full ngEHT reference array 1 as described in \citet{2023Roelofs} and \citet{2023Doeleman}. The 22 mm VLBI sites will yield a very dense $(u, v)$ coverage and image reconstructions with much better dynamic ranges.
Since the GRRT images of our exotic models were ray-traced only in total intensity, we focus only on the Stokes~$\mathcal{I}$ data here.

For the dilaton models, a noticeable feature is the structural and total flux variability excesses of the SANE model over the MAD model. This feature can be used to easily distinguish the two $b_*=0.504$, $R_\mathrm{high} = 40$, $i_\mathrm{los} = \ang{20}$, $\theta_\mathrm{PA}=0$ models through the Stokes~$\mathcal{I}$ measurements. Typically MAD simulations are accompanied by more violent outflow eruptions and are thus overall more variable than SANE models, at least for the standard models considered in this work. As the nonrotating dilaton models stay below $\phi_\mathrm{mag} = 10$, no strong eruptions occur. The variability excess of the SANE dilaton model will be investigated in a future study.

The two SANE $a_*=0.25$, $R_\mathrm{high} = 80$, \m87 Kerr-Newman models in \Cref{fig:synthetic_data_dkn} differ in their charge $q_*$ parameters -- one model has $q_* = 0$ and the other has $q_*=0.9$. Both models are similarly compact as the standard SANE models shown in \Cref{fig:synthetic_data_m87}, which also have $R_\mathrm{high} = 80$.
Given that the size of the event horizon is $1 + \sqrt{1 - a_*^2 -q_*^2}$, the size of the photon ring is noticeably smaller for the $q_*=0.9$ model.
The $q_*=0$ model shows some foreground emission with a low surface brightness inside of the photon ring. For the $q_*=0.9$ model, there is a bright ring of foreground emission inside of the photon ring.
As a result, the black hole shadow appears to be substantially smaller. The smaller ring size caused by the black hole charge is clearly identifiable in the shift of the total intensity visibility minima location toward longer baseline distances.

Building upon earlier horizon-scale GRMHD-GRRT variability studies of \m87 by \citet{2022Satapathy} and \sgra by \citet{2022Georgiev}, we analyze the influence of model variability on parameter inference, taking the full forward modeling chain for the synthetic data generation into account for the first time. We note here that telescope-based gain errors and inaccuracies in the signal stabilization method impact closure quantities when data are averaged in time or frequency.
\Cref{fig:m87_cphase_vs_time} shows closure phase variability from three standard \m87 models on the ALMA-LMT-SMT and LMT-PV-SMT triangles, which show little variability in the observational data over the six-day-long extend of the 2017 EHT observations \citep{2022Satapathy}.
Different realizations that are based on the same model image have comparatively little influence on the \m87 closure phase variability for the two triangles analyzed here. 
The dominant source of variability comes from the differences between individual model frames, which is most evident for the ALMA-LMT-SMT data of the SANE, $a_*=-0.94$, $R_\mathrm{high}=160$ model.

The LMT-PV-SMT data do not have discriminative power for the three \m87 models shown.
With only a few observations, the ALMA-LMT-SMT data alone cannot be used to distinguish between the models either. With long-term monitoring, the model variability can be used to distinguish between the example parameters here; the one SANE model differs substantially from the two MADs based on the overall degree of variability.
The MAD models themselves differ only by the black hole spin. Their degree of variability is comparable, but the median closure phases value, measured over years, is lower for the $a_*=0$ model compared to the $a_*=0.94$ one.

These simple examples make use of only a small part of the simulation model parameter space and features in the data. In \citet{zingularity2}, we study how well a Bayesian neural network can distinguish standard models when the full parameter space and all data products are taken into account. We find that our neural network, by using the full information content of the data, is able to distinguish all model parameters in the presence of the specific closure phase degeneracies presented here. Furthermore, the network is robust against the intrinsic model variations -- a network trained only on the first half of GRRT frames can accurately predict the parameters of the latter half of frames for each model.

\Cref{fig:sgra_cphase_reals} shows ALMA-LMT-SMT and ALMA-SPT-SMA closure phase variations for three synthetic data realizations of the MAD $a_*=0.5$, $R_\mathrm{high}=160$, $i_\mathrm{los}=\ang{30}$, $\theta_\mathrm{PA}=0$, standard \sgra model for the 7 April 2017 EHT track.
Due to the short gravitational timescale of \sgra, multiple GRRT frames make up a single synthetic dataset, resulting in significant differences concerning how well different realizations of the same model agree with the observational data.
This effect is most evident in the ALMA-LMT-SMT data. For ALMA-SPT-SMA, we found that only a few model frames agree with the observational measurements after 12 UT.
Consequently, the MAD $a_*=0.5$, $R_\mathrm{high}=160$, $i_\mathrm{los}=\ang{30}$ model was found to pass the \ac{eht} data constraints used in \citet{eht-SgrAv}, when variability is not taken into account.
In \citet{zingularity2}, we show that a suitably trained Bayesian deep neural network can accurately fit parameters of interest, even when different model realizations exhibit significant variations in specific data products.

\section{Summary and conclusions}
\label{sec:synthdataconclusions}

In preparation for machine-learning-based \ac{grmhd}-\ac{grrt} parameter inference with \ac{eht} data \citep{zingularity2}, we produced a comprehensive synthetic data library matching different millimeter \ac{vlbi} observations. Most of our 962,000 datasets are made to match the 2017 \ac{eht} observations of \sgra and \m87 on 7 and 11 April, respectively.
Additionally, we studied synthetic data from possible future observations of the EHT+AMT and EHT+ngEHT.
Next to a standard set of Kerr black hole models, that broadly sample the $\phi_\mathrm{mag}$, $a_*$, $R_\mathrm{high}$, and $i_\mathrm{los}$ parameter space, we also included Kerr-Newman and EMDA gravity black hole solutions in our library.

We presented a substantial upgrade in the \ac{casa}-based \ac{eht} data reduction pathway, obtaining the hitherto highest quality \ac{eht} data. The calibration process with its upgrades presented here is taken into account for our generation of realistic synthetic data. Thereby, direct comparisons of the raw visibilities can be made between the observational and synthetic data.

We analyzed the synthetic data from several selected models for dedicated case studies. We also demonstrated that a long-term monitoring of \m87 is needed to discriminate between different models through closure phase measurements. For the  $a_*=-0.5$ and $R_\mathrm{high} = 80$ \m87 model, we found that intrinsic model variability has a strong imprint on the total intensity visibility data, while variations in polarized light are mostly caused by data corruption effects. For the $a_*=0$, $R_\mathrm{high} = 1$, $i_\mathrm{los} = \ang{90}$, $\theta_\mathrm{PA}=0$, \sgra model on the other hand, the linear polarization phases are well suited for distinguishing between $\phi_\mathrm{mag}$ accretion states.
We showed that the MAD $a_*=0.5$, $R_\mathrm{high}=160$, $i_\mathrm{los}=\ang{30}$, $\theta_\mathrm{PA}=0$, \sgra model, which was found to pass the stationary \ac{eht} data constraints considered in \citet{eht-SgrAv}, does not actually fit the measured \ac{aa}-\ac{sp}-\ac{sm} closure phase.
It is clear that different models have different salient features that can be used as discriminating factors.

For our \sgra dilaton models, we found the intrinsic model variability in total intensity within a single \ac{eht} observing track to be a good indicator for $\phi_\mathrm{mag}$. Surprisingly, the variability is higher for SANE models, opposite to the typical behavior of the standard \ac{grmhd}-\ac{grrt} models.
For the SANE $a_*=0.25$, $R_\mathrm{high} = 80$, Kerr-Newmann \m87 model, the possible presence of a large enough black hole charge is easily identifiable by the shrinking black hole shadow size.

The case studies presented in this work are useful to obtain intuition for the \ac{eht} observational and model data.
Equipped with the gained insights, we have performed parameter inference studies with machine learning methods in \citet{zingularity2} and \citet{zingularity3}.

\begin{acknowledgements}

We thank the anonymous referee for their insight and helpful
suggestions that have improved this paper.

We thank Feryal \"Ozel and Dimitrios Psaltis for their help in setting up the synthetic data generation pipeline using CyVerse and the Open Science Grid. We thank Illyoung Choi for optimizing the access to the CyVerse data storage for our workflow and for his many swift fixes of issues.

This publication is part of the M2FINDERS project which has received funding from the European Research Council (ERC) under the European Union's Horizon 2020 Research and Innovation Programme (grant agreement No 101018682).

JD is supported by NASA through the NASA Hubble Fellowship grant HST-HF2-51552.001A, awarded by the Space Telescope Science Institute, which is operated by the Association of Universities for Research in Astronomy, Incorporated, under NASA contract NAS5-26555.

JR received financial support for this research from the International Max Planck Research School (IMPRS) for Astronomy and Astrophysics at the Universities of Bonn and Cologne. 
This material is based upon work supported by the National Science Foundation under Award Numbers DBI-0735191, DBI-1265383, and DBI-1743442. URL: \url{www.cyverse.org}.
This research was done using services provided by the OSG Consortium \citep{osg07, osg09, osgdoi1, osgdoi2}, which is supported by the National Science Foundation awards \#2030508 and \#1836650.
This research used the Pegasus Workflow Management Software funded by the National Science Foundation under grant \#1664162.
Computations were performed on the HPC system Cobra at the Max Planck Computing and Data Facility, as well as on the Iboga and Calea clusters at the ITP Frankfurt.

MW is supported by a Ram\'on y Cajal grant RYC2023-042988-I from the Spanish Ministry of Science and Innovation.

This research made use of the high-performance computing Raven-GPU cluster of the Max Planck Computing and Data Facility.

\end{acknowledgements}

\bibliographystyle{aa} 
\bibliography{link_to_init}

\begin{appendix}
\onecolumn
\section{Synthetic data versus ground truth} \label{app:SvG}
\begin{figure*}[h!]
    \centering
    \includegraphics[height=5.8cm]{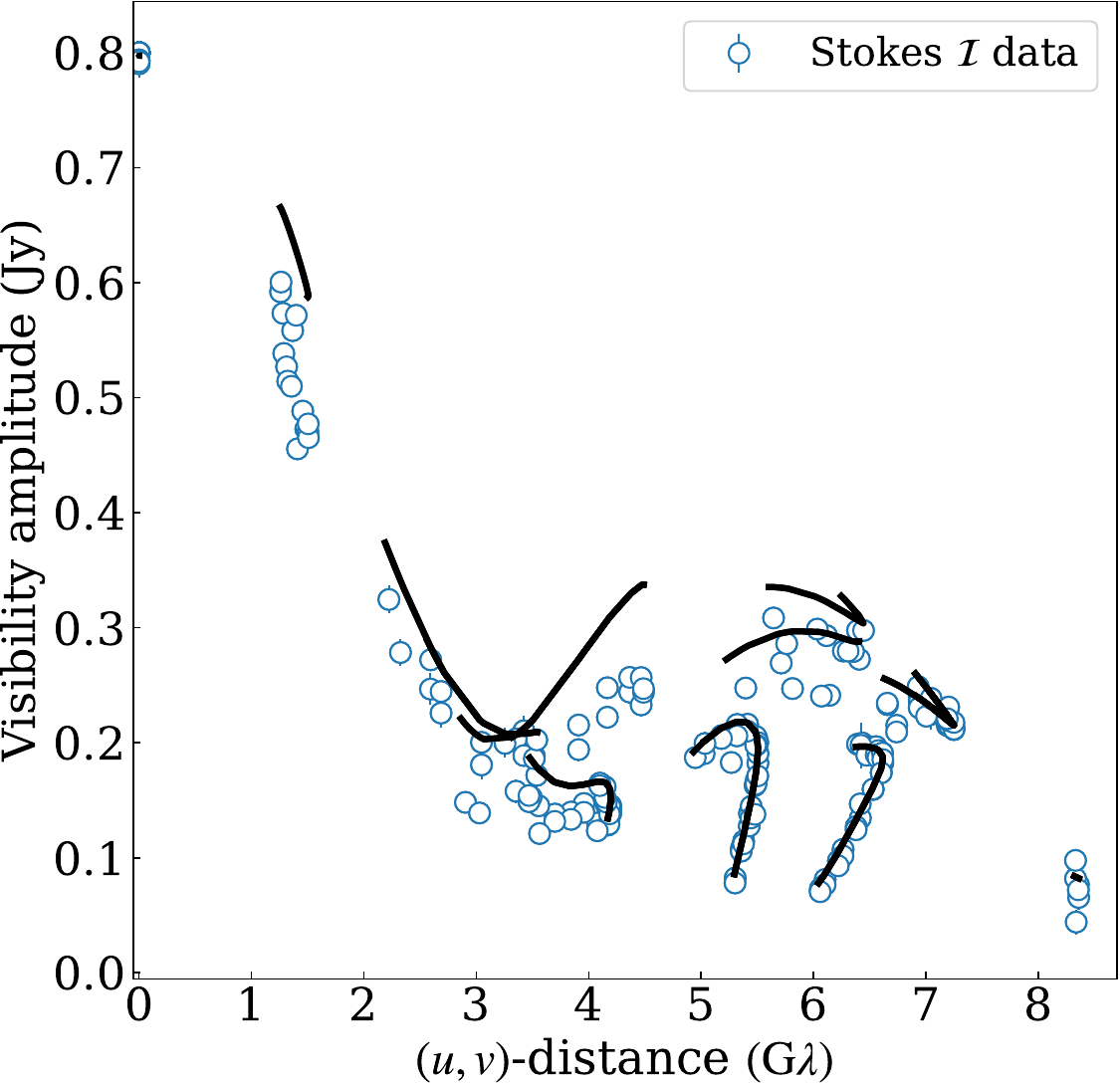}
    \includegraphics[height=5.8cm]{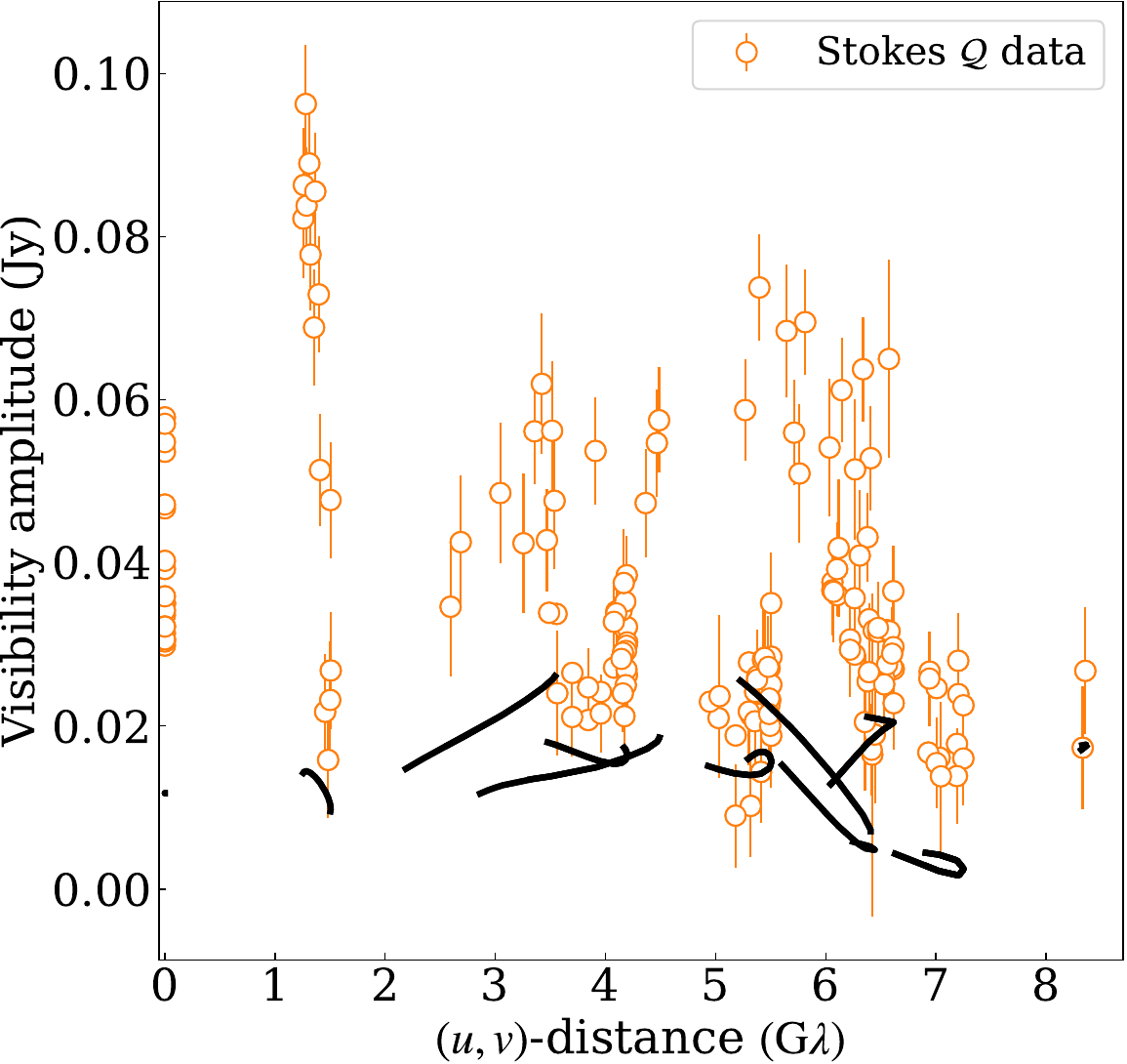}
    \includegraphics[height=5.8cm]{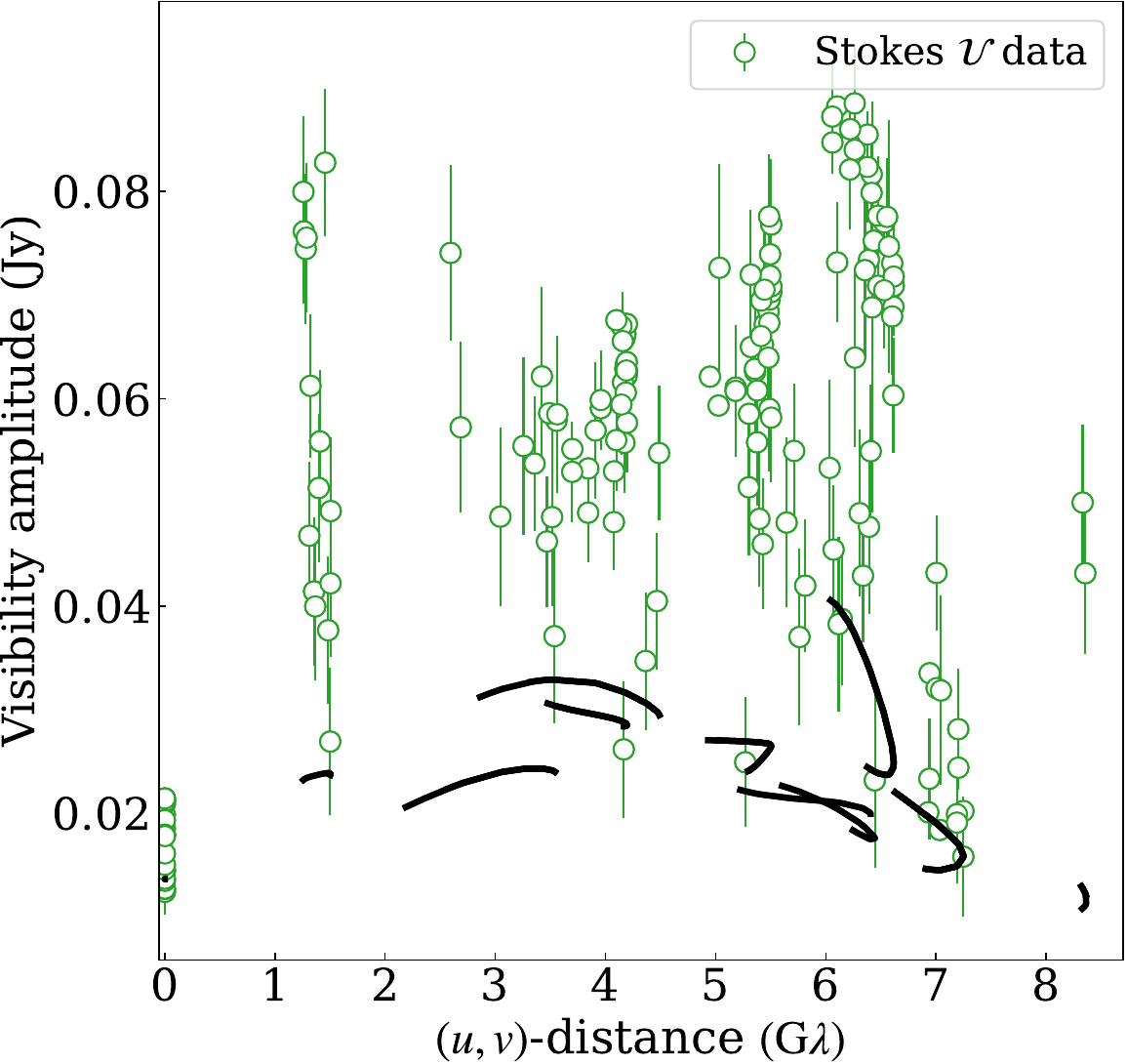}
    \includegraphics[height=5.8cm]{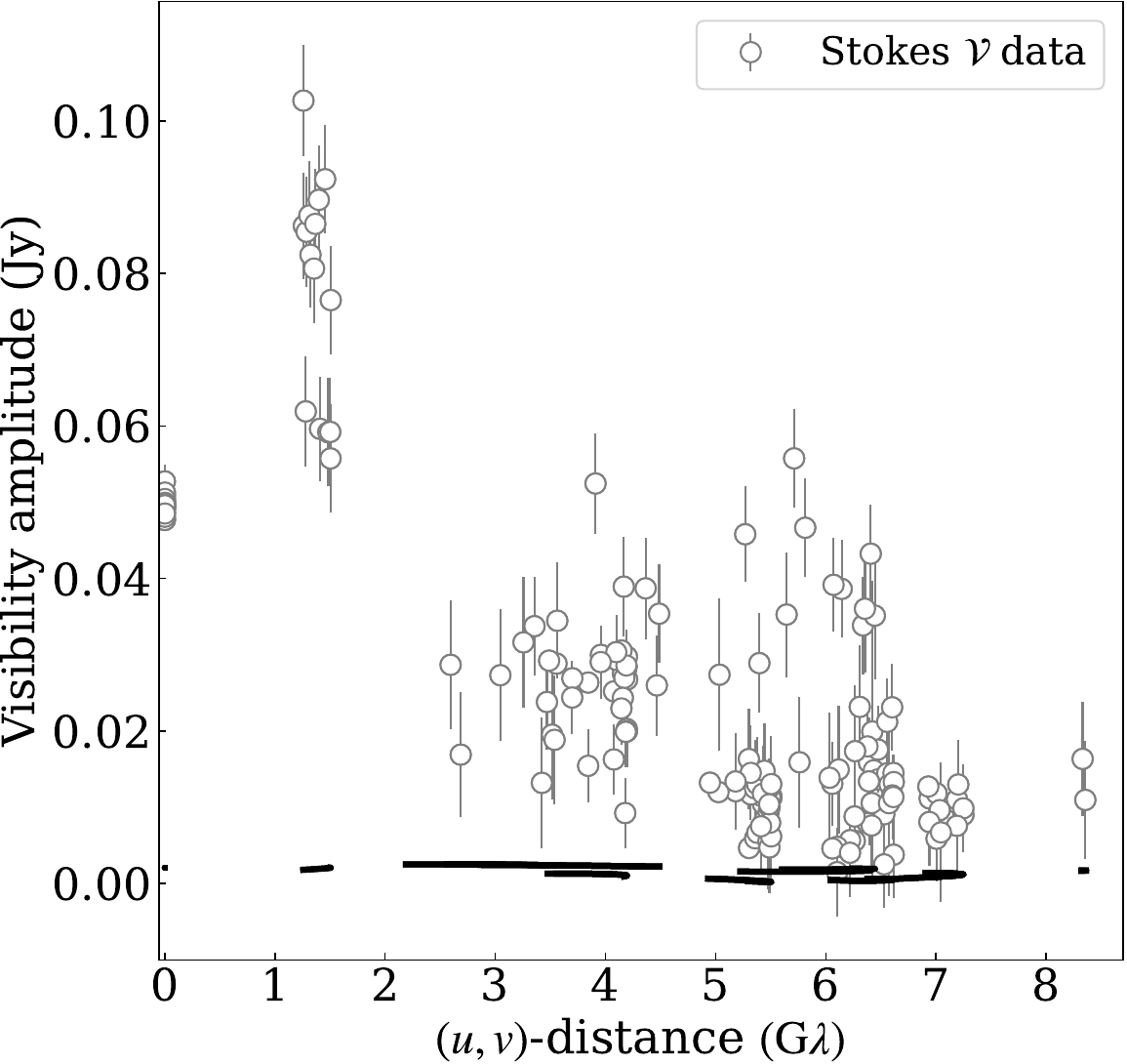}
    \includegraphics[width=12.5cm]{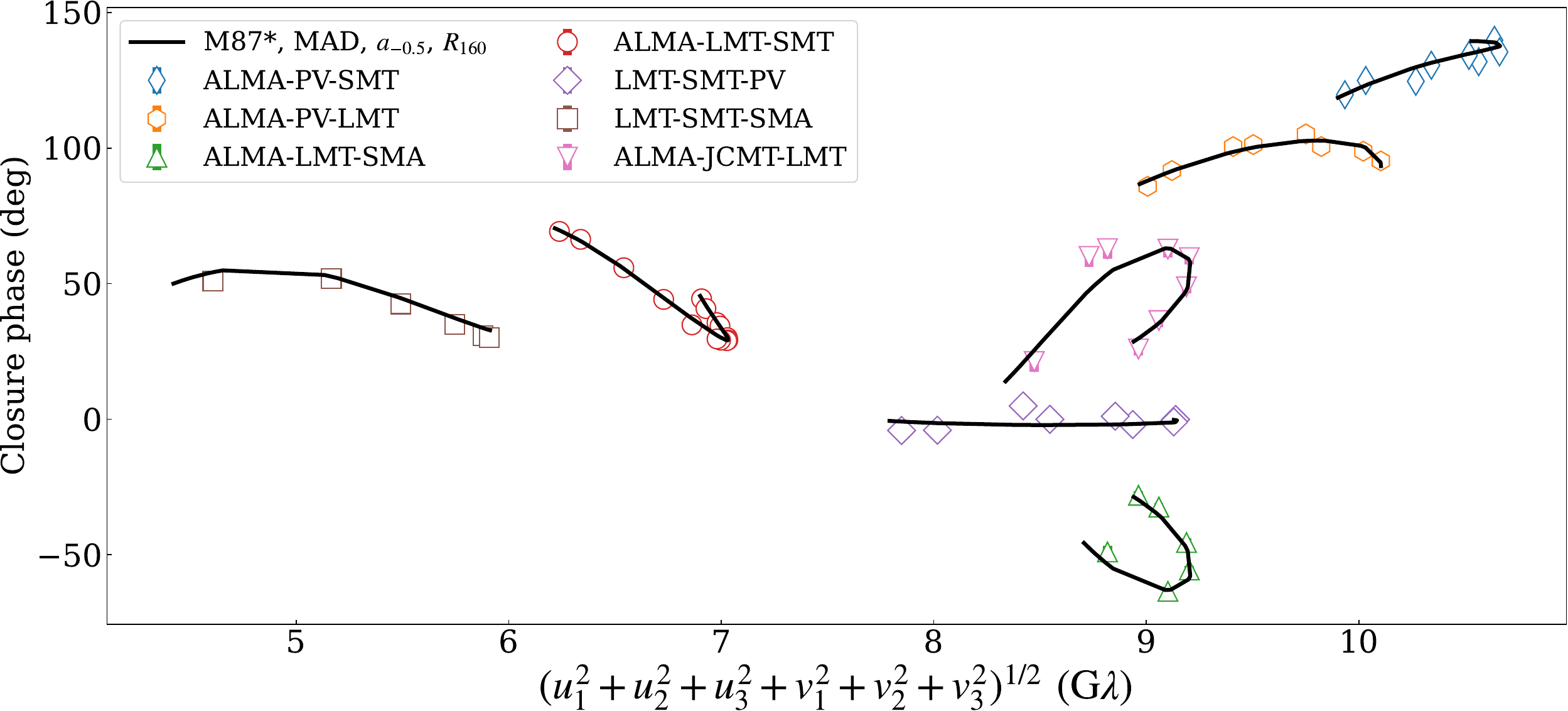}
    \caption{Direct comparison of synthetic- and corresponding model data from the \m87 MAD image of \Cref{fig:synthetic_data_m87}. The four top panels show amplitudes of the different Stokes parameters being primarily affected by RCP, LCP gain errors and $\mathcal{D}$-terms. The closure phases shown in the bottom panel are mostly unaffected by the errors along the VLBI signal path. The synthetic data are averaged over VLBI scan durations. For most Stokes $\mathcal{I}$ and closure phase data points, the displayed thermal noise error bars are smaller than the plotted symbols. The model data is shown with a black line.}
    \label{fig:m87_SvG}
\end{figure*}

\FloatBarrier

In \Cref{fig:m87_SvG}, we show how the various simulated errors along the signal path affect the measured data, compared to model data produced by just an FFT as it would be measured by a perfect instrument.
The Stokes~$\mathcal{I}$ data is primarily affected by the the gross telescope gain errors $\mathcal{G}_\mathrm{err}$ that affect both polarizations in the same way, as well as antenna pointing errors. These errors are responsible for a loss of measured signal amplitude on several baselines.
Due to polarization leakage, artificial source polarization is created, causing the linear polarization synthetic data amplitude to exceed the model.
Circular polarization is very weak in the source and the produced signal is almost entirely the result of R$-$L telescope gain errors.
The only baseline-dependent effect simulated by \symba is phase decoherence from atmospheric fluctuations. As those are well calibrated by \rpicard, the simulated closure phases show no substantial deviations from the model.
The analysis of the \sgra synthetic data output gives similar results, albeit with additional noise due to the intrinsic source variability and the interstellar scattering screen.

\end{appendix}
\end{document}